\documentclass[prd,preprint,superscriptaddress,nofootinbib,%
tightenlines]{revtex4}
\usepackage{mathrsfs}
\usepackage{latexsym,bm}
\usepackage{graphicx}
\usepackage{indentfirst}
\usepackage{slashed}
\usepackage{amssymb}
\usepackage{amsmath}
\usepackage{bbm}
\usepackage{color}
\usepackage[dvipsnames]{xcolor}
\usepackage{epsfig}
\usepackage[titletoc]{appendix}
\usepackage{multirow}%
\usepackage{rotating}
\usepackage{epstopdf}
\usepackage{extarrows}
\usepackage{tabularx}
\usepackage{soul} 
\usepackage{xcolor}
\usepackage{lipsum}

\usepackage{makecell}

\usepackage{caption}
\usepackage{subcaption}
\usepackage[colorlinks=true,allcolors=BlueViolet,hyperfootnotes=false]{hyperref}

\usepackage[utf8]{inputenc}
\usepackage{slashed}

\newcommand{\be}{\begin{equation}} \newcommand{\ee}{\end{equation}}
\newcommand{\ba}{\begin{array}{c}} \newcommand{\ea}{\end{array}}
\newcommand{\bea}{\begin{eqnarray}} \newcommand{\eea}{\end{eqnarray}}

\begin{document}
\title{ Decay rates of $\Lambda _b^0 \to \Lambda _c^ + {\ell ^ - }{\bar \nu _\ell }$ using helicity analysis and phase-moment parametrization}
\author{Sara~Rahmani}
\email{s.rahmani120@gmail.com}
\affiliation{School of Physics and Electronics, Hunan University, Changsha 410082, China}
\begin{abstract}
Based on the helicity method, formulae for the semileptonic transition of $\Lambda _b^0 \to \Lambda _c^ + {\ell ^ - }{\bar \nu _\ell }$ including lepton mass effects are derived. In order to calculate the form factors of the $\Lambda_b$ baryon transition matrix element, we employ the phase-moment parameterization and perform fits to the Lattice QCD data. With the help of the obtained form factors, six helicity amplitudes and the differential decay widths are evaluated. Through appropriate angular integrations, we express the helicity flip, helicity nonflip integrated decay rates, and the lepton-side forward-backward asymmetry. We present a numerical analysis of these physical observables. We obtain the mentioned physical quantities by performing fits to the Lattice QCD data using the well-known Boyd-Grinstein-Lebed parametrization. Comparisons with other experimental and theoretical data are also discussed.
  
\end{abstract}

\maketitle

\section{Introduction}
Semileptonic decays of heavy baryons are important physical procedures for the determination of the Cabibbo-Kobayashi-Maskawa (CKM) matrix elements, i.e. $|V_{cb}|$ in transitions of $b \to c$, investigation of the $CP$ violation in the Standard Model (SM), and for the new physics signal. In the SM, these decays are mediated by $W^\pm$ bosons with universal coupling to leptons. The ratio $\mathcal{R}(\Lambda _c^ + ) \equiv BR(\Lambda _b^0 \to \Lambda _c^ + {\tau ^ - }{\bar \nu _\tau })/BR(\Lambda _b^0 \to \Lambda _c^ + {\mu ^ - }{\bar \nu _\mu })$ is another important quantity to test the SM and lepton flavour universality and provides a sensitive probe of SM extensions. LHCb Collaboration observed the semileptonic $b$-baryon decay $\Lambda _b^0 \to \Lambda _c^ + {\tau ^ - }{\bar \nu _\tau }$
with a significance of 6.1 $\sigma$ \cite{LHCb:2022piu}. They reported the branching fraction of the tauonic channel $BR(\Lambda _b^0 \to \Lambda _c^ + {\tau ^ - }{\bar \nu _\tau }) = (1.50 \pm 0.16 \pm 0.25 \pm 0.23)\% $, and derived $\mathcal{R}(\Lambda _c^ + ) \equiv BR(\Lambda _b^0 \to \Lambda _c^ + {\tau ^ - }{\bar \nu _\tau })/BR(\Lambda _b^0 \to \Lambda _c^ + {\mu ^ - }{\bar \nu _\mu }) = 0.242 \pm 0.026 \pm 0.040 \pm 0.059$, where the last term uncertainty comes from the branching fraction of muonic channel \cite{LHCb:2022piu}. They measured the shape of the differential decay rate $d\Gamma/dq^2$ and the associated Isgur-Wise function for $\Lambda _b^0 \to \Lambda _c^ + {\mu ^ - }{\bar \nu _\mu }$, where their results were compared with expectations from
Heavy-Quark Effective Theory (HQET) and lattice QCD (LQCD) predictions \cite{LHCb:2017vhq}.
\par
There has been a significant number of theoretical papers on the weak decay of $\Lambda _b^0 \to \Lambda _c^ + {\ell ^ - }{\bar \nu _\ell }$, where $\ell = e, \mu, \tau$. Some of the phenomenological frameworks used are the light-cone sum rule (LCSR) \cite{Duan:2022uzm, Miao:2022bga}, covariant light-front quark model \cite{Zhu:2018jet, Li:2021qod}, QCD
sum rules \cite{Azizi:2018axf, Zhao:2020mod}, the light-front
approach under the diquark picture \cite{Zhao:2018zcb}, and Isgur-Wise functions in the heavy quark limit \cite{Huang:2005mea,Bernlochner:2018bfn, Rahmani:2020kjd,Hassanabadi:2014cla,Rahmani:2016wgu}.
Generally, LQCD is the most reliable method to compute the hadronic form factors. The LQCD computations had been performed by Detmold {\it{et. al.}} for $\Lambda _b^{} \to \Lambda _c^{}{\ell ^ - }{{\bar \nu }_\ell }$ form factors \cite{Detmold:2015aaa}, in which the form factor results to
the physical pion mass and the continuum limit had been extrapolated.  
\par
In a broader context, three has been numerous discussions regarding weak decays involving $b \to c$ in their quark level. From the experimental point of view, the CDF Collaboration presented the measurements of the quantities $\frac{{BR(\Lambda _b^0 \to \Lambda _c^ + {\mu ^ - }{{\bar \nu }_\mu })}}{{BR(\Lambda _b^0 \to \Lambda _c^ + {\pi ^ - })}}$, and also reported the branching fractions of $\Lambda _b^0 \to \Lambda _c^{}{(2595)^ + }{\mu ^ - }{{\bar \nu }_\mu },\Lambda _b^0 \to \Lambda _c^{}{(2625)^ + }{\mu ^ - }{{\bar \nu }_\mu },\Lambda _b^0 \to \Sigma _c^{}{(2455)^0}{\pi ^ + }{\mu ^ - }{{\bar \nu }_\mu },$ and $\Lambda _b^0 \to \Sigma _c^{}{(2455)^{ +  + }}{\pi ^ - }{\mu ^ - }{{\bar \nu }_\mu }$
relative to the branching fraction of the ${\Lambda _b^0 \to \Lambda _c^ + {\mu ^ - }{{\bar \nu }_\mu }}$ decay \cite{CDF:2008hqh}. From a theoretical perspective, Duan {\it{et. al.}} presented the theoretical framework for the form factors of $\Lambda _b^0 \to \Lambda _c^{}{(2595)^ + }$ within the light-cone QCD sum rules recently \cite{Duan:2024lnw}. Lu {\it{et. al.}} investigated the weak decay ${\Sigma _b} \to \Sigma _c^*\ell {{\bar \nu }_\ell }$, which is a specific form of ${\frac{1}{2}^ + } \to {\frac{3}{2}^ + }$ transition amplitude using the light-front quark model, and derived the hadronic form factors \cite{Lu:2023rmq}.
The widths and polarization asymmetries of semileptonic decay ${\Xi _b} \to \Xi _c^{(')}\ell {{\bar \nu }_\ell }$ have been computed by Ke {\it{et. al.}}, and they have discussed the effect of the mixing angle on the ratios of the mentioned weak decays \cite{Ke:2024aux}.
The decay amplitudes of bottomed baryons and mesons including $b \to c$ in their transition have been probed via the light diquarks picture \cite{Zhang:2024ery}.
\par
The research of weak decay processes with transitions of bottomed to charmed baryons will enhance our comprehension of the dynamics of these bottom baryon decays.
The study of semileptonic $b$ baryon modes provides insights into the flavour sector. Besides, the helicity analysis allows one to provide the evaluation of the angular decay distributions effectively. We aim to use this formalism to discuss weak decay rates of $\Lambda_b^0 \to \Lambda_c^+ \ell \bar{\nu}_\ell$.
This work is organized as follows: in section \ref{section: formalism} we will introduce the effective Lagrangian for the weak induced $b \to c$ transitions. We give an analytical analysis of the helicity amplitudes and twofold angular distributions. The Phase-Moment (PM) form factors are also defined. The results and discussions are collected in section \ref{section: numeric}. Finally, a conclusion will be given in section \ref{section: conclusion}.

\section{Formalism}
\label{section: formalism}
\subsection{Invariants and Helicity amplitudes}
\label{section: helicity}
\begin{figure}[h]
    \includegraphics[width=0.5\textwidth]{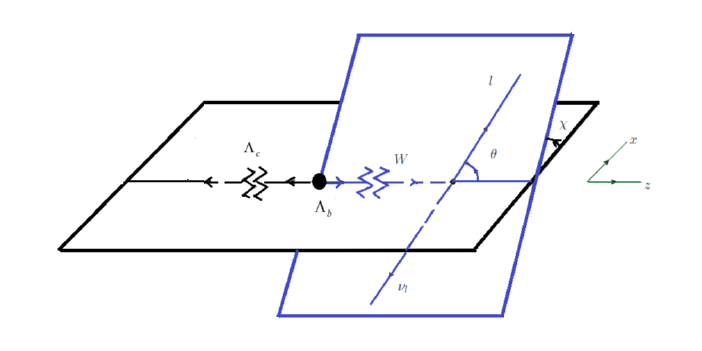}
    \caption{Definition of the polar $\theta$ and azimuthal $\chi $ angles.}
    \label{fig: weak decay angles}
\end{figure}

The effective Fermi Hamiltonian depicted corresponding semileptonic $b$ to $c$ induced transitions reads as
\begin{equation}
    \mathcal{H}_{eff} = \frac{{{G_F}}}{{\sqrt 2 }}{V_{cb}}\bar \ell{\gamma ^\mu }(1 - {\gamma _5}){\nu _\ell}\bar c{\gamma _\mu }(1 - {\gamma _5})b,
\end{equation}
in which, we investigate the semileptonic decay of $\Lambda _b^0$ to $\Lambda _c^ +$, where the masses and momenta are denoted by ${B_1}({p_1},{M_1}) \to {B_2}({p_2},{M_2}) + \ell({p_\ell},{m_\ell}) + {\nu _\ell}({p_\nu },0)$. In the SM, the corresponding invariant matrix element for semileptonic decay of $\Lambda _b^0$ to $\Lambda _c^ +$ can be written as
\begin{equation}
    \begin{array}{l}
{\cal M}{\rm{ = }}\frac{{{G_F}}}{{\sqrt 2 }}{V_{cb}}\left\langle {{\Lambda _c}|\bar c{{\cal O}_\mu }b|{\Lambda _b}} \right\rangle {\ell^ - }{{\cal O}^\mu }{\nu _\ell}\\
{\rm{ = }}\frac{{{G_F}}}{{\sqrt 2 }}{V_{cb}}{H_\mu }{L^\mu },
\end{array}
\end{equation}
where ${{\cal O}_\mu } = {\gamma _\mu }(1 - {\gamma _5})$, and $G_F$ is the Fermi coupling constant. Hence, the hadronic matrix elements of vector and axial vector currents $J^{V, A}_\mu$ for this process are usually parameterized as 
\begin{equation}
    \begin{array}{l}
{\cal M}_\mu ^V = \left\langle {{B_2}|\bar c{\gamma _\mu }b|{B_1}} \right\rangle  = {{\bar u}_2}({p_2})\left[ {F_1^V({q^2}){\gamma _\mu } - \frac{{F_2^V({q^2})}}{{{M_1}}}i{\sigma _{\mu \nu }}{q^\nu } + \frac{{F_3^V({q^2})}}{{{M_1}}}{q_\mu }} \right]{u_1}({p_1}),\\
{\cal M}_\mu ^A = \left\langle {{B_2}|\bar c{\gamma _\mu }{\gamma _5}b|{B_1}} \right\rangle  = {{\bar u}_2}({p_2})\left[ {F_1^A({q^2}){\gamma _\mu } - \frac{{F_2^A({q^2})}}{{{M_1}}}i{\sigma _{\mu \nu }}{q^\nu } + \frac{{F_3^A({q^2})}}{{{M_1}}}{q_\mu }} \right]{\gamma _5}{u_1}({p_1}),
\end{array}
\end{equation}
in terms of six dimensionless invariant form factors $F_i^{V,A}, i = {1, 2, 3}$ which are a function of the momentum transfer squared ($q^2$). We take ${\sigma _{\mu \nu }} = \frac{i}{2}({\gamma _\mu }{\gamma _\nu } - {\gamma _\nu }{\gamma _\mu })$, and also $q = {p_1} - {p_2}$. Gamma matrices have been considered to have ${\gamma _5} = i{\gamma ^0}{\gamma _1}{\gamma _2}{\gamma _3} =  - \left( {\begin{array}{*{20}{c}}
0&1\\
1&0
\end{array}} \right)$. 
We study two particles with spin $\frac{1}{2}$ which leads to $J^P = (0^+, 1^-)$ and $J^P = (0^-, 1^+)$ for vector and axial-vector currents, respectively. ${\lambda _2} =  \pm \frac{1}{2}$ are defined as the helicity components of the daughter baryon. $J=1, 0$ are the two angular momentum of the rest frame $W_{off-shell}$. Thus the helicity components of the $W$ are ${\lambda _W} =  \pm 1,0$ and ${\lambda _W} = t$ for $J=1$ and $J=0$ respectively. In fact, whenever we write ${\lambda_{W}} = t$, this has to be considered as ${\lambda_{W}} = 0$ with $(J = 0)$. One has ${\lambda _1} = {\lambda _2} - {\lambda _W}$ due to the angular momentum conservation. 
Note that we work on ${B_1} \to {B_2} + W_{off-shell}$, where $W^-_{off-shell} \to {\ell ^ - }{{\bar \nu }_\ell }$. In the $B_1 (\Lambda _b^0)$ rest frame, we take the four momentums of the $B_2 (\Lambda _c^ + )$ and virtual $W$ boson along the negative $z$ and $+z$ directions respectively (see Fig. \ref{fig: weak decay angles}). Hence, in the $B_1$ rest frame, the four-vector polarization $\epsilon^\mu({\lambda _W})$ are \cite{Kadeer:2005aq}
\begin{equation}
    \begin{gathered}
  \epsilon^\mu ( \pm 1) = \frac{1}{{\sqrt 2 }}\left( {0, \pm 1, - i,0} \right), \hfill \\
  {\epsilon^\mu }(0) = \frac{1}{{\sqrt {{q^2}} }}\left( {|{{\vec p}_2}|,0,0, - {q_0}} \right), \hfill \\
  {{\epsilon}^\mu }(t) = \frac{1}{{\sqrt {{q^2}} }}\left( {{q_0},0,0, - |{{\vec p}_2}|} \right), \hfill \\ 
\end{gathered} 
\end{equation}
where the four-momentum of $W_{off-shell}$ is
\begin{equation}
    {q^\mu } = ({q_0},0,0,|{\vec p_2}|),
\end{equation}
with $q_0$ and $p_2$ are the energy and momentum of $W$ in the $B_1$ rest frame, defined by

\begin{equation}
    \begin{gathered}
  {q_0} = \frac{1}{{2{M_1}}}(M_1^2 - M_2^2 + {q^2}), \hfill \\
  |{p_2}{|^2} = E_2^2 - M_2^2, \hfill \\
  {E_2} = \frac{1}{{2{M_1}}}(M_1^2 + M_2^2 - {q^2}). \hfill \\ 
    \end{gathered} 
\end{equation} 

Further, the four-momentum of $B_1$ and $B_2$ are read as
\begin{equation}
    \begin{gathered}
  p_1^\mu  = ({M_1},0,0,0), \hfill \\
  p_2^\mu  = ({E_2},0,0, - |{{\vec p}_2}|). \hfill \\ 
\end{gathered} 
\end{equation}

The baryon spinors are then given by \cite{Auvil:1966eao}
\begin{equation}
    \begin{gathered}
  {{\bar u}_2}( \pm \frac{1}{2},{p_2}) = \sqrt {{E_2} + {M_2}} \left( {\chi _ \pm ^{\text{\dag }},\frac{{ \mp |{{\vec p}_2}|}}{{{E_2} + {M_2}}}\chi _ \pm ^{\text{\dag }}} \right), \hfill \\
  {u_1}( \pm \frac{1}{2},{p_1}) = \sqrt {2{M_1}} \left( \begin{gathered}
  {\chi _ \pm } \hfill \\
  0 \hfill \\ 
\end{gathered}  \right), \hfill \\ 
\end{gathered} 
\end{equation}
where ${\chi _ + } = \left( \begin{gathered}
  1 \hfill \\
  0 \hfill \\ 
\end{gathered}  \right),{\chi _ - } = \left( \begin{gathered}
  0 \hfill \\
  1 \hfill \\ 
\end{gathered}  \right)$ are the usual Pauli two-spinors. To calculate the branching ratios and forward-backward asymmetries, which are physical observables, the helicity amplitudes are applicable. We analyze the decay by means of helicity amplitudes
\begin{equation}
    H_{{\lambda _2},{\lambda _W}}^{V/A} = \mathcal{M}_\mu ^{V/A}({\lambda _2}){\epsilon}^{{\text{\dag }}\mu }({\lambda _W}).
\end{equation}
Therefore, we obtain the following relations for vector and axial vector helicity amplitudes
\begin{equation}
\begin{gathered}
    H_{\frac{1}{2},0}^V = \frac{{\sqrt {{Q_ - }} }}{{\sqrt {{q^2}} }}(({M_1} + M_2)F_1^V + \frac{{{q^2}}}{{{M_1}}}F_2^V),\\
    H_{\frac{1}{2},1}^V =  - \sqrt {2{Q_ - }} (F_1^V + \frac{{{M_1} + {M_2}}}{{{M_1}}}F_2^V),\\
    H_{\frac{1}{2},t}^V = \sqrt {\frac{{{Q_ + }}}{{{q^2}}}} (({M_1} - {M_2})F_1^V + \frac{{{q^2}}}{{{M_1}}}F_3^V),\\
    H_{\frac{1}{2},0}^A = \sqrt {\frac{{{Q_ + }}}{{{q^2}}}} (({M_2} - {M_1})F_1^A + \frac{{{q^2}}}{{{M_1}}}F_2^A),\\
    H_{\frac{1}{2},1}^A = \sqrt {2{Q_ + }} (F_1^A + \frac{{{M_2} - {M_1}}}{{{M_1}}}F_2^A),\\
    H_{\frac{1}{2},t}^A = \sqrt {\frac{{{Q_ - }}}{{{q^2}}}} ( - ({M_1} + {M_2})F_1^A + \frac{{{q^2}}}{{{M_1}}}F_3^A),\\
\end{gathered}
\end{equation}
with the abbreviations ${Q_ \pm } = {({M_1} \pm {M_2})^2} - {q^2}$. With the help of parity relations, one has the remaining helicity amplitudes
\begin{equation}
    \begin{gathered}
  H_{ - \frac{1}{2},0}^A =  - H_{\frac{1}{2},0}^A,H_{ - \frac{1}{2}, - 1}^A =  - H_{\frac{1}{2},1}^A,H_{ - \frac{1}{2},t}^A =  - H_{\frac{1}{2},t}^A, \hfill \\
  H_{ - \frac{1}{2}, - 1}^V = H_{\frac{1}{2},1}^V,H_{ - \frac{1}{2},0}^V = H_{\frac{1}{2},0}^V,H_{ - \frac{1}{2},t}^V = H_{\frac{1}{2},t}^V. \hfill \\ 
\end{gathered} 
\end{equation}
\subsection{Twofold angular distribution}
The components of helicity amplitudes have been separated into vector and axial-vector parts. The rate and the angular decay distributions will be expressed with the $V-A$ current in terms of total helicity amplitudes of the current-induced transition as
\begin{equation}
    {H_{{\lambda _2},{\lambda _W}}} = H_{{\lambda _2},{\lambda _W}}^V - H_{{\lambda _2},{\lambda _W}}^A,
\end{equation}
where ${\lambda _W} = 0$, ${\lambda _W} =  \pm 1$, and ${\lambda _W} = t$
are referred to as longitudinal (spin 1), transverse (spin 1), and (spin 0) time components of the off-shell $W$-boson respectively. The definitions of the helicity structure functions with their parity properties are shown in Table \ref{Table: helicity}. 

\begin{table}[htb]
\caption{Helicity structure functions \cite{Gutsche:2015mxa}.}
\label{Table: helicity}
\begin{tabularx}{1\textwidth}{>{\centering\arraybackslash}X |>{\centering\arraybackslash}X}
\hline\hline

Parity-Conserving & Parity-violating
\\
\hline
${\mathcal{H}_U} = |{H_{\frac{1}{2}, + }}{|^2} + |{H_{ - \frac{1}{2}, - }}{|^2}$ & ${\mathcal{H}_P} = |{H_{\frac{1}{2}, + }}{|^2} - |{H_{ - \frac{1}{2}, - }}{|^2}$
\\
${\mathcal{H}_L} = |{H_{\frac{1}{2},0}}{|^2} + |{H_{ - \frac{1}{2},0}}{|^2}$ & ${\mathcal{H}_{L_P}} = |{H_{\frac{1}{2}, 0 }}{|^2} - |{H_{ - \frac{1}{2}, 0 }}{|^2}$
\\
${\mathcal{H}_S} = |{H_{\frac{1}{2},t}}{|^2} + |{H_{ - \frac{1}{2},t}}{|^2}$ & ${\mathcal{H}_{S_P}} = |{H_{\frac{1}{2}, t }}{|^2} - |{H_{ - \frac{1}{2}, t }}{|^2}$
\\
${\mathcal{H}_{LT}} = \operatorname{Re} ({H_{\frac{1}{2},+}}H_{- \frac{1}{2},0}^{\text{\dag }} + {H_{  \frac{1}{2},0}}H_{ - \frac{1}{2},-}^{\text{\dag }})$ & ${\mathcal{H}_{{LT}_P}} = \operatorname{Re} ({H_{\frac{1}{2},+}}H_{- \frac{1}{2},0}^{\text{\dag }} - {H_{  \frac{1}{2},0}}H_{ - \frac{1}{2},-}^{\text{\dag }})$
\\
${\mathcal{H}_{ST}} = \operatorname{Re} ({H_{\frac{1}{2},+}}H_{- \frac{1}{2},t}^{\text{\dag }} + {H_{  \frac{1}{2},t}}H_{ - \frac{1}{2},-}^{\text{\dag }})$ & ${\mathcal{H}_{{ST}_P}} = \operatorname{Re} ({H_{\frac{1}{2},+}}H_{- \frac{1}{2},t}^{\text{\dag }} - {H_{  \frac{1}{2},t}}H_{ - \frac{1}{2},-}^{\text{\dag }})$
\\
${\mathcal{H}_{SL}} = \operatorname{Re} ({H_{\frac{1}{2},0}}H_{ \frac{1}{2},t}^{\text{\dag }} + {H_{-  \frac{1}{2},0}}H_{ - \frac{1}{2},t}^{\text{\dag }})$ & ${\mathcal{H}_{{SL}_P}} = \operatorname{Re} ({H_{\frac{1}{2},0}}H_{ \frac{1}{2},t}^{\text{\dag }} - {H_{-  \frac{1}{2},0}}H_{ - \frac{1}{2},t}^{\text{\dag }})$
\\
\hline\hline
\end{tabularx}
\end{table}

\par
The differential rate is obtained from the twofold angular-dependent decay distribution with regard to $q^2$ and the polar angle $\theta$. We have \cite{Gutsche:2015mxa}
\begin{equation}
    \begin{gathered}
  \frac{{{d^2}\Gamma ({\Lambda _b} \to {\Lambda _c}{l^ + }{\nu _l})}}{{d{q^2}d\cos \theta }} 
   = \frac{{G_F^2|{V_{cb}}{|^2}}}{{{{(2\pi )}^4}}}\frac{{|{\vec p_2}|}}{{128M_1^2}}\left( {1 - \frac{{m_\ell^2}}{{{q^2}}}} \right){H_{\mu \nu }}{L^{\mu \nu }}, \hfill \\ 
\end{gathered} 
\end{equation}
where $m_\ell$ is lepton mass, and $|{\vec p_2}| = {\lambda ^{1/2}}(M_1^2,M_2^2,{q^2})/2{M_1}$ is the momentum of the daughter baryon in terms of the Källén function $\lambda (x,y,z) \equiv {x^2} + {y^2} + {z^2} - 2(xy + yz + zx)$. Equivalently, this quantity can also be defined as $|{\vec p_2}| = \frac{{\sqrt {{Q_ + }{Q_ - }} }}{{2{M_1}}}$. The Lorentz contraction of hadronic and leptonic tensors can be written in terms of the helicity amplitudes using orthonormal and complete helicity basis ($\epsilon^\mu (\lambda_W)$). They have the orthonormality property and completeness relations as
\begin{equation}
    {g_{{\lambda _W},{{\lambda '}_W}}} = \epsilon_\mu ^{{\dag }}({\lambda _W})\epsilon^\mu ({\lambda '_W}),\hfill \\
    {g_{\mu \nu }} = \sum\limits_{{\lambda _W},{\lambda '_{W}}} {{\epsilon_\mu }({\lambda _W}){\epsilon}_\nu ^{\text{\dag }}({\lambda '_W}){g_{{\lambda _W},{{\lambda '}_W}}}}, 
\end{equation}
with ${\lambda _W},{\lambda '_W} = t, \pm ,0$ and ${g_{{\lambda _W},{{\lambda '}_W}}} = {\text{diag}}( + , - , - , - )$. Therefore, one can rewrite the contraction of leptonic and hadronic tensors as

\begin{equation}
    \begin{gathered}
  {L^{\mu \nu }}{H_{\mu \nu }} =  
  {L_{\mu '\nu '}}{g^{\mu '\mu }}{g^{\nu '\nu }}{H_{\mu \nu }} \hfill \\
   = \sum\limits_{{\lambda _W},{{\lambda '}_W},{{\lambda ''}_W},{{\lambda '''}_W}} {{L_{\mu '\nu '}}{\epsilon}^{\mu '}({\lambda _W}){\epsilon}_{}^{{\text{\dag }}\mu }({{\lambda ''}_W}){g_{{\lambda _W},{{\lambda ''}_W}}}{\epsilon}^{{\text{\dag }}\nu '}({{\lambda '}_W}){\epsilon}_{}^\nu ({{\lambda '''}_W}){g_{{{\lambda '}_W},{{\lambda '''}_W}}}{H_{\mu \nu }}}     \hfill \\
     \equiv \sum\limits_{{\lambda _W},{{\lambda '}_W}} {L({\lambda _W},{{\lambda '}_W})} {g_{{\lambda _W},{\lambda _W}}}{g_{{{\lambda '}_W},{{\lambda '}_W}}}H({\lambda _W},{{\lambda '}_W}), \hfill \\ 
\end{gathered} 
\end{equation}
with the leptonic and hadronic tensors,
\begin{equation}
\label{Eq: leptonic tensor}
L({\lambda _W},{\lambda '_W}) = \epsilon^{\mu '}({\lambda _W})\epsilon^{{\text{\dag }}\nu '}({\lambda '_W}){L_{\mu '\nu '}},\    
\end{equation}
and 
\begin{equation}
H({\lambda _W},{\lambda '_W}) = {\epsilon}^{{\text{\dag }}\mu }({\lambda _W}){\epsilon}^\nu ({\lambda '_W}){H_{\mu \nu }}.
\end{equation}
The hadronic tensor for the $\Lambda _b^0 \to \Lambda _c^ + $ transition is then evaluated in connection with helicity amplitudes by
\begin{equation}
    {H_{\mu \nu }} = {({\mathcal{M}^V} - {\mathcal{M}^A})_\mu }({\mathcal{M}^V} - {\mathcal{M}^A})_\nu ^{\text{\dag }}.
\end{equation}
On the other hand, we can write the hadronic tensor in terms of bilinear products of the helicity amplitudes,
\begin{equation}
\mathcal{ H} ({\lambda _W},{\lambda '_W}) \equiv \sum\limits_{{\lambda _2}} {{H_{{\lambda _2}{\lambda _W}}}H_{{\lambda _2}{{\lambda '}_W}}^{\text{\dag }}}, 
\end{equation}
in the $\Lambda _b^0$ rest frame. Therefore, we can write
\begin{equation}
    \mathcal{H}({\lambda _W},{\lambda '_W}) = \left( {\begin{array}{*{20}{c}}
  \mathcal{H}_S&{{H_{\frac{1}{2},t}}H_{\frac{1}{2}, + }^{\text{\dag }}}&\mathcal{H}_{SL}&{{H_{ - \frac{1}{2},t}}H_{ - \frac{1}{2}, - }^{\text{\dag }}} \\ 
  {{H_{\frac{1}{2}, + }}H_{\frac{1}{2},t}^{\text{\dag }}}&{|{H_{\frac{1}{2}, + }}{|^2}}&{{H_{\frac{1}{2}, + }}H_{\frac{1}{2},0}^{\text{\dag }}}&0 \\ 
  \mathcal{H}_{SL}&{{H_{\frac{1}{2},0}}H_{\frac{1}{2}, + }^{\text{\dag }}}&\mathcal{H}_{L}&{{H_{ - \frac{1}{2},0}}H_{ - \frac{1}{2}, - }^{\text{\dag }}} \\ 
  {{H_{ - \frac{1}{2}, - }}H_{ - \frac{1}{2},t}^{\text{\dag }}}&0&{{H_{ - \frac{1}{2}, - }}H_{ - \frac{1}{2},0}^{\text{\dag }}}&{|{H_{ - \frac{1}{2}, - }}{|^2}} 
\end{array}} \right).
\end{equation}
Leptonic tensor in the helicity space  in the $W$ rest frame can be given by
\begin{equation}
    {L_{\mu \nu }} = Tr[({\slashed{k}_1} + {m_l}){\mathcal{O}_\mu }{\slashed{k}_2}{\mathcal{O}_\nu }],
\end{equation}
for ${W^ - } \to {\ell ^ - }{\bar \nu _\ell }$, and
\begin{equation}
    {L_{\mu \nu }} = Tr[({\slashed{k}_1} - {m_l}){\mathcal{O}_\nu }{\slashed{k}_2}{\mathcal{O}_\mu }],
\end{equation}
for the case of ${W^ + } \to {\ell ^ + }{ \nu _\ell }$. We reach at
\begin{equation}
    {L_{\mu \nu }} = 8(k_1^\mu k_2^\nu  + k_1^\nu k_2^\mu  - {k_1}.{k_2}{g^{\mu \nu }} \pm i{\varepsilon ^{\mu \nu \alpha \beta }}{k_{1\alpha }}{k_{2\beta }}),
\end{equation}
where the upper/lower sign refers to the two ${\ell^ - }{\bar \nu _\ell}/{\ell^ + }{\nu _\ell}$ configurations. The Levi-Civita tensor in Minkowski space is taken as ${\varepsilon _{0123}} =  + 1$. We shall work in the $W$ rest frame. The explicit expressions of $q$, $k_1$ and $k_2$, the four momentums of $W$ virtual particle, charged lepton and neutrino respectively can be represented as
\begin{equation}
    \begin{gathered}
  {q^\mu } = \left( {\sqrt {{q^2}} ,0,0,0} \right), \hfill \\
  k_1^\mu  = \left( {{E_1},|\vec{k_1}|\sin \theta \cos \chi ,|\vec{k_1}|\sin \theta \sin \chi ,|\vec{k_1}|\cos \theta } \right), \hfill \\
  k_2^\mu  = \left( {|\vec{k_1}|, - |\vec{k_1}|\sin \theta \cos \chi , - |\vec{k_1}|\sin \theta \sin \chi , - |\vec{k_1}|\cos \theta } \right), \hfill \\ 
\end{gathered} 
\end{equation}
where the energy and three-momentum of the charged lepton are
\begin{equation}
    \begin{gathered}
  {E_1} = \frac{{{q^2} + m_\ell^2}}{{2\sqrt {{q^2}} }}, \hfill \\
  |{k_1}| = \frac{{{q^2} - m_\ell^2}}{{2\sqrt {{q^2}} }}. \hfill \\ 
\end{gathered} 
\end{equation}
The polarization vectors in this rest frame are then given by
\begin{equation}
    \begin{gathered}
  \epsilon^\mu ( \pm 1) = \frac{1}{{\sqrt 2 }}\left( {0, \mp 1, - i,0} \right), \hfill \\
  \epsilon^\mu (0) = \left( {0,0,0,1} \right), \hfill \\
  \epsilon^\mu (t) = \left( {1,0,0,0} \right). \hfill \\ 
\end{gathered} 
\end{equation}
Using polarization vectors as well as Eq. (\ref{Eq: leptonic tensor}), we obtain
\begin{equation}
\label{Eq: leptonic matrix}
    \begin{gathered}
  {(2{q^2}v)^{ - 1}}L({\lambda _W},{{\lambda '}_W})(\theta ,\chi ) \hfill \\
   = \left( {\begin{array}{*{20}{c}}
  0&0&0&0 \\ 
  0&{{{(1 \mp \cos \theta )}^2}}&{ \mp \frac{2}{{\sqrt 2 }}(1 \mp \cos \theta )\sin \theta {e^{i\chi }}}&{{{\sin }^2}\theta {e^{2i\chi }}} \\ 
  0&{ \mp \frac{2}{{\sqrt 2 }}(1 \mp \cos \theta )\sin \theta {e^{ - i\chi }}}&{2{{\sin }^2}\theta }&{ \mp \frac{2}{{\sqrt 2 }}(1 \pm \cos \theta )\sin \theta {e^{i\chi }}} \\ 
  0&{{{\sin }^2}\theta {e^{ - 2i\chi }}}&{ \mp \frac{2}{{\sqrt 2 }}(1 \pm \cos \theta )\sin \theta {e^{ - i\chi }}}&{{{(1 \pm \cos \theta )}^2}} 
\end{array}} \right) \hfill \\
   + {\delta _\ell}\left( {\begin{array}{*{20}{c}}
  4&{ - \frac{4}{{\sqrt 2 }}\sin \theta {e^{ - i\chi }}}&{4\cos \theta }&{\frac{4}{{\sqrt 2 }}\sin \theta {e^{i\chi }}} \\ 
  { - \frac{4}{{\sqrt 2 }}\sin \theta {e^{i\chi }}}&{2{{\sin }^2}\theta }&{ - \frac{2}{{\sqrt 2 }}\sin 2\theta {e^{i\chi }}}&{ - 2{{\sin }^2}\theta {e^{2i\chi }}} \\ 
  {4\cos \theta }&{ - \frac{2}{{\sqrt 2 }}\sin 2\theta {e^{ - i\chi }}}&{4{{\cos }^2}\theta }&{\frac{2}{{\sqrt 2 }}\sin 2\theta {e^{i\chi }}} \\ 
  {\frac{4}{{\sqrt 2 }}\sin \theta {e^{ - i\chi }}}&{ - 2{{\sin }^2}\theta {e^{ - 2i\chi }}}&{\frac{2}{{\sqrt 2 }}\sin 2\theta {e^{ - i\chi }}}&{2{{\sin }^2}\theta } 
\end{array}} \right), \hfill \\ 
\end{gathered} 
\end{equation}
where velocity-type and helicity-flip factors are defined by
\begin{equation}
    v = 1 - m_\ell^2/{q^2},{\delta _\ell} = m_\ell^2/2{q^2},
\end{equation}
and the upper/lower sign in the nonflip part of Eq. (\ref{Eq: leptonic matrix}) corresponds to the two configurations ${\ell^ - }{\bar \nu _\ell}/{\ell^ + }{\nu _\ell}$. To obtain the polar angle distribution, one has to integrate the above matrix over the azimuthal angle $\chi$ as
\begin{equation}
    \begin{gathered}
  {\frac{1}{{2\pi }}(2{q^2}v)^{ - 1}}L({\lambda _W},{{\lambda '}_W})(\theta ) \hfill \\
   = \left( {\begin{array}{*{20}{c}}
  0&0&0&0 \\ 
  0&{{{(1 \mp \cos \theta )}^2}}&0&0 \\ 
  0&0&{2{{\sin }^2}\theta }&0 \\ 
  0&0&0&{{{(1 \pm \cos \theta )}^2}} 
\end{array}} \right) \hfill \\
   + {\delta _\ell }\left( {\begin{array}{*{20}{c}}
  4&0&{4\cos \theta }&0 \\ 
  0&{2{{\sin }^2}\theta }&0&0 \\ 
  {4\cos \theta }&0&{4{{\cos }^2}\theta }&0 \\ 
  0&0&0&{2{{\sin }^2}\theta } 
\end{array}} \right). \hfill \\ 
\end{gathered} 
\end{equation}
We take the lower sign in the matrix to evaluate ${W^ + } \to {\ell ^ + }{\nu _\ell }$. In the helicity space, one has the contraction of hadronic and leptonic tensors as follows:
\begin{equation}
    \begin{gathered}
  {L^{\mu \nu }}{H_{\mu \nu }(\theta)} = (4{q^2}v\pi) \{ (1 + {\cos ^2}\theta )\mathcal{ H} _U + 2{\sin ^2}\theta \mathcal{ H} _L + 2\cos \theta \mathcal{ H} _P \hfill \\
   + 2{\delta _\ell}[{\sin ^2}\theta \mathcal{ H} _U + 2{\cos ^2}\theta \mathcal{ H} _L + 2\mathcal{ H} _S - 4\cos \theta \mathcal{ H} _{SL}\}.  \hfill \\ 
\end{gathered} 
\end{equation}
Thus, the twofold polar angular distribution becomes
\begin{equation}
\label{eq: twofold}
    \begin{gathered}
  \frac{{{d}\Gamma ({\Lambda _b} \to {\Lambda _c}{l^ + }{\nu _\ell})}}{{d{q^2}d\cos \theta }} = \frac{{G_F^2|{V_{cb}}{|^2}|{p_2}|{q^2}{v^2}}}{{64{{(2\pi )}^3}M_1^2}} \times \{ (1 + {\cos ^2}\theta ){\mathcal{ H}_U} + 2{\sin ^2}\theta {\mathcal{ H}_L} + 2\cos \theta {\mathcal{ H}_P} \hfill \\
   + 2{\delta _\ell}[{\sin ^2}\theta {\mathcal{ H}_U} + 2{\cos ^2}\theta \mathcal{ H}_L + 2\mathcal{ H}_S - 4\cos \theta \mathcal{ H}_{SL}\}.  \hfill \\ 
\end{gathered} 
\end{equation}
The first three terms in the above equation arise from lepton helicity nonflip contributions, while the latter four terms are proportional to $\delta_\ell$, the lepton helicity flip. Finally, the distribution regarding transfer momentum $q^2$, will be
\begin{equation}
\label{eq: gamma to q2}
    \frac{{d\Gamma ({\Lambda _b} \to {\Lambda _c}{l^ + }{\nu _l})}}{{d{q^2}}} = \frac{{G_F^2|{V_{cb}}{|^2}|{p_2}|{q^2}{v^2}}}{{24{{(2\pi )}^3}M_1^2}} \times {\mathcal{H}_{tot}},
\end{equation}
by integrating over cos$\theta$ in Eq. (\ref{eq: twofold}) and considering 
\begin{equation}
    {\mathcal{H}_{tot}} = {\mathcal{H}_U} + {\mathcal{H}_L} + {\delta _\ell}[{\mathcal{H}_U} + {\mathcal{H}_L} + 3{\mathcal{H}_S}].
\end{equation}
In Eq. (\ref{eq: gamma to q2}), helicity functions $\mathcal{H}_U, \mathcal{H}_L, \mathcal{H}_S$  characterize the effect of the
hadronic structure and are defined by the momentum transfer squared $q^2$ varying in the range  $m_\ell ^2 \leqslant {q^2} \leqslant {({M_{\Lambda_b}} - {M_{\Lambda_c})^2}}$. The partial helicity width $\Gamma_I$ corresponds to one of the combinations of differential helicity amplitudes from Table \ref{Table: helicity} which are defined as
\begin{equation}
        \frac{{d\Gamma_I(nf) }}{{d{q^2}}} = \frac{{G_F^2|{V_{cb}}{|^2}|{p_2}|{q^2}{v^2}}}{{24{{(2\pi )}^3}M_1^2}}  {\mathcal{H}_{I}},
\label{Eq: nf}
\end{equation}
for the helicity nonflip ($nf$) structure functions where $I = U, L, P, L_P, LT, {LT}_P$, and
\begin{equation}
        \frac{{d{\tilde \Gamma }_I(hf) }}{{d{q^2}}} = {\delta _\ell }\frac{{G_F^2|{V_{cb}}{|^2}|{p_2}|{q^2}{v^2}}}{{24{{(2\pi )}^3}M_1^2}}  {\mathcal{H}_{I}},
\label{Eq: hf}
\end{equation}
for the helicity flip ($hf$) structure functions with $I = U$, $L$, $LT$, $P$, $SL$, $S$, $L_P$, $S_P$, ${SL}_P$, ${LT}_P$, ${ST}_P$.
\par
Further, to investigate another important physical observable, we can define the forward-backward asymmetry for the charged leptons by
\begin{equation}
    \mathcal{A}^\ell_{FB} (q^2) = \frac{{\int\limits_0^1 {d\cos \theta \frac{{d\Gamma }}{{d{q^2}d\cos \theta }}}  - \int\limits_{ - 1}^0 {d\cos \theta \frac{{d\Gamma }}{{d{q^2}d\cos \theta }}} }}{{\int\limits_0^1 {d\cos \theta \frac{{d\Gamma }}{{d{q^2}d\cos \theta }}}  + \int\limits_{ - 1}^0 {d\cos \theta \frac{{d\Gamma }}{{d{q^2}d\cos \theta }}} }}
     =  - \frac{3}{4}\frac{{{\mathcal{H}_P} + 4{\delta _\ell}{\mathcal{H}_{SL}}}}{{{\mathcal{H}_{tot}}}},
\end{equation}
in which we have considered ${W^ - } \to {\ell^ - }{\nu _\ell}$ in the extraction of leptonic and hadronic tensors. The “forward” and “backward” regions include $\theta  \in [0,\pi /2]$ and $\theta  \in [\pi /2,\pi]$ respectively.

\subsection{The transition form factors with the phase-moment parameterization}
The issue key in studying the decay widths of weak decay processes is the determination of the form factors which offers valuable insights into the internal dynamics of hadrons. Yao {\it{et. al.}} proposed a new parameterization for the form factors of $b\to c$ transitions \cite{Yao:2019vty}. This phase-moment parametrization (PM) reads as
\begin{equation}
\mathcal{F}({q^2}) = \mathcal{F}({t_0})\prod\limits_{n = 0}^\infty  {\exp \left[ {\frac{{{q^2} - {t_0}}}{{{t_ + }}}\mathcal{A}_n^\mathcal{F}({t_0}){{\left( {\frac{{{q^2}}}{{{t_ + }}}} \right)}^n}} \right]}, 
\end{equation}
where the phase moments are given by
\begin{equation}
    \mathcal{A}_n^\mathcal{F}({s_0}) \equiv \frac{1}{\pi }\int\limits_{{s_ + }}^\infty  {\frac{{ds'}}{{s' - {s_0}}}} \frac{{\psi (s')}}{{{{(s'/{s_ + })}^{n + 1}}}}.
\end{equation}
In our process, $\mathcal{F}$ is any of the vector and axial vector form factors, defined in helicity amplitudes in section \ref{section: helicity}. For example, we take the vector form factors as follows:
\begin{equation}
    F_i^V({q^2}) = F_i^V({t_0})\exp \left[ {\frac{{{q^2} - {t_0}}}{{{t_ + }}}\mathcal{A}_0^{F_i^V} + \frac{{{q^2} - {t_0}}}{{{t_ + }}}\mathcal{A}_1^{F_i^V}\frac{{{q^2}}}{{{t_ + }}} + ...} \right],i \in \{ 1,2,3\}, 
\end{equation}
where $t_+ = (M_1 + M_2)^2$, and $t_0$ is the subtraction point which can be any value in the range of $[0, t_-]$ with  $t_- = (M_1 - M_2)^2$. $\mathcal{A}_n^\mathcal{F}$ are called phase moments since they are related to the phases of the form factors in the scattering region of $\Lambda_b^0 \Lambda_c^+$.  Various functional forms including an exponential shape have been parametrized for helicity form factors of $\Lambda_b^0 \to \Lambda_c^+ \mu^- \bar{\nu_\mu}$ by LHCb \cite{LHCb:2017vhq}. The exponential type of form factors was also studied in different quark potential models within Isgur-Wise function approaches, through the transitions of $\Lambda_b^0 \to \Lambda_c^+$ \cite{Rahmani:2020kjd}. Besides, Huang {\it{et. al.}} assumed an exponential form to parameterize the baryonic Isgur-Wise function within HQET using the sum rule analysis \cite{Huang:2005mea}.
\section{Numerical Analysis}
\label{section: numeric}
\begin{figure}
     \centering
     \begin{subfigure}[b]{0.49\textwidth}
         \centering
         \includegraphics[width=\textwidth]{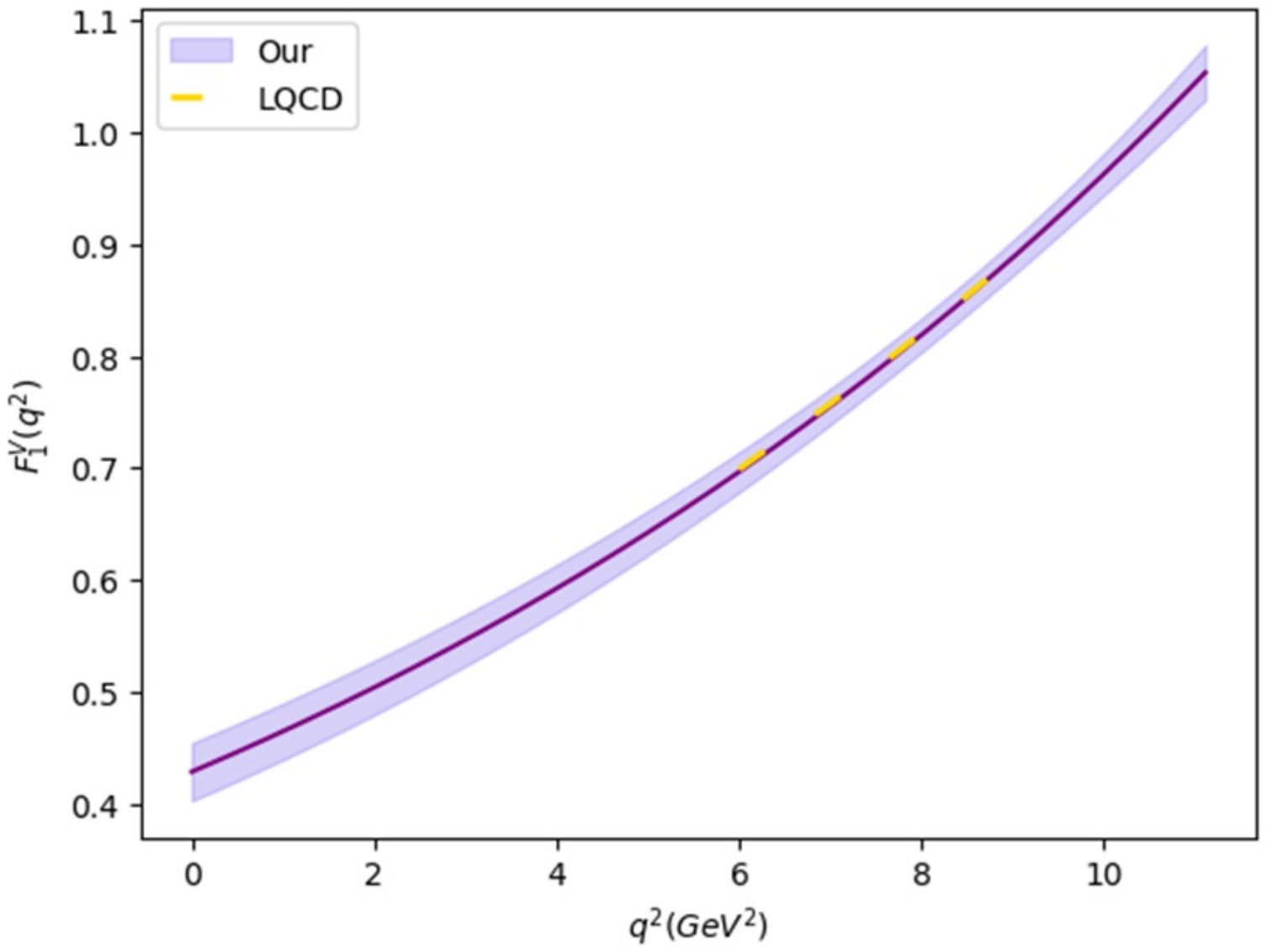}
     \end{subfigure}
     \hfill
     \begin{subfigure}[b]{0.49\textwidth}
         \centering
         \includegraphics[width=\textwidth]{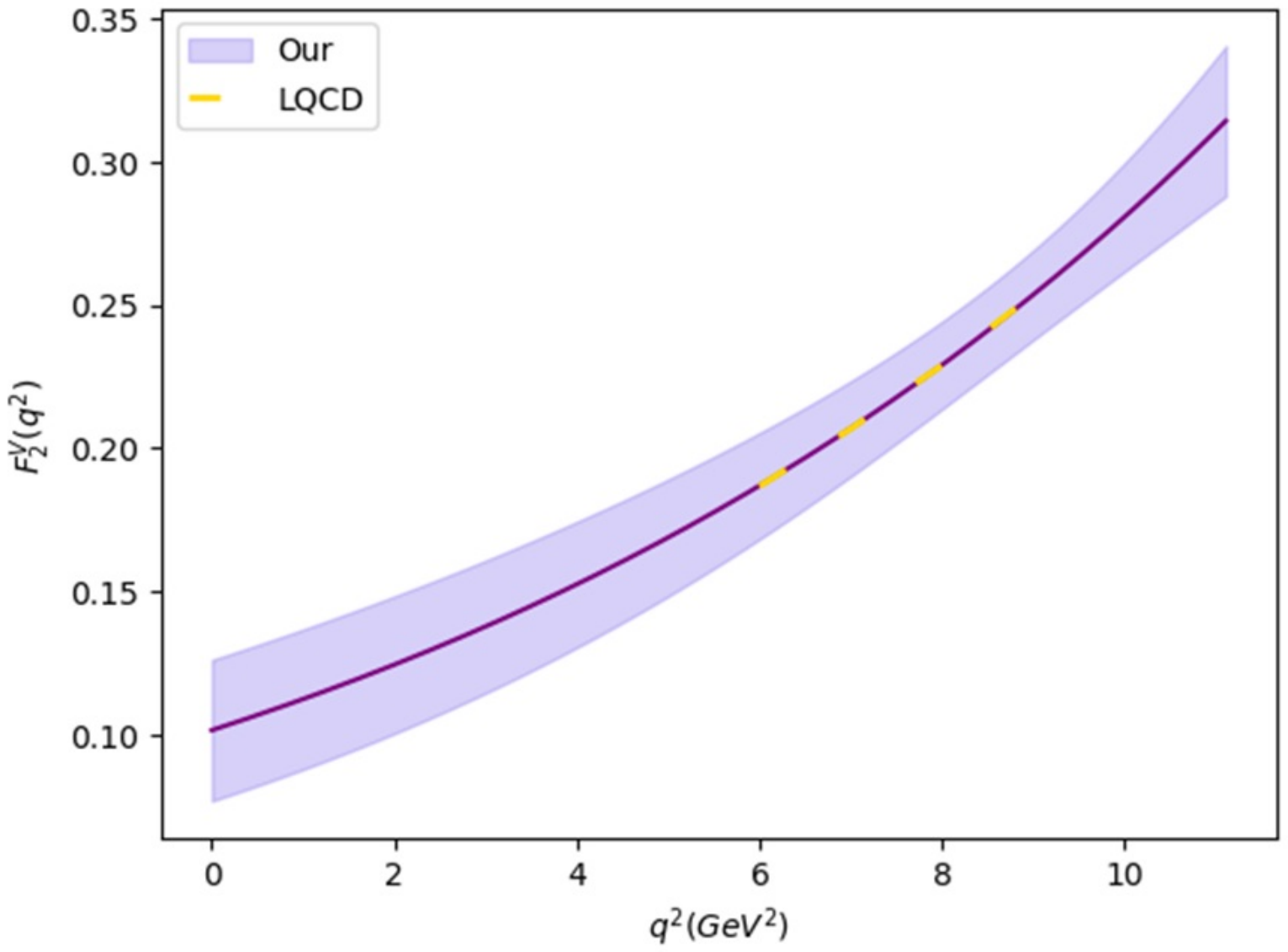}
     \end{subfigure}
     \hfill
     \begin{subfigure}[b]{0.49\textwidth}
         \centering
         \includegraphics[width=\textwidth]{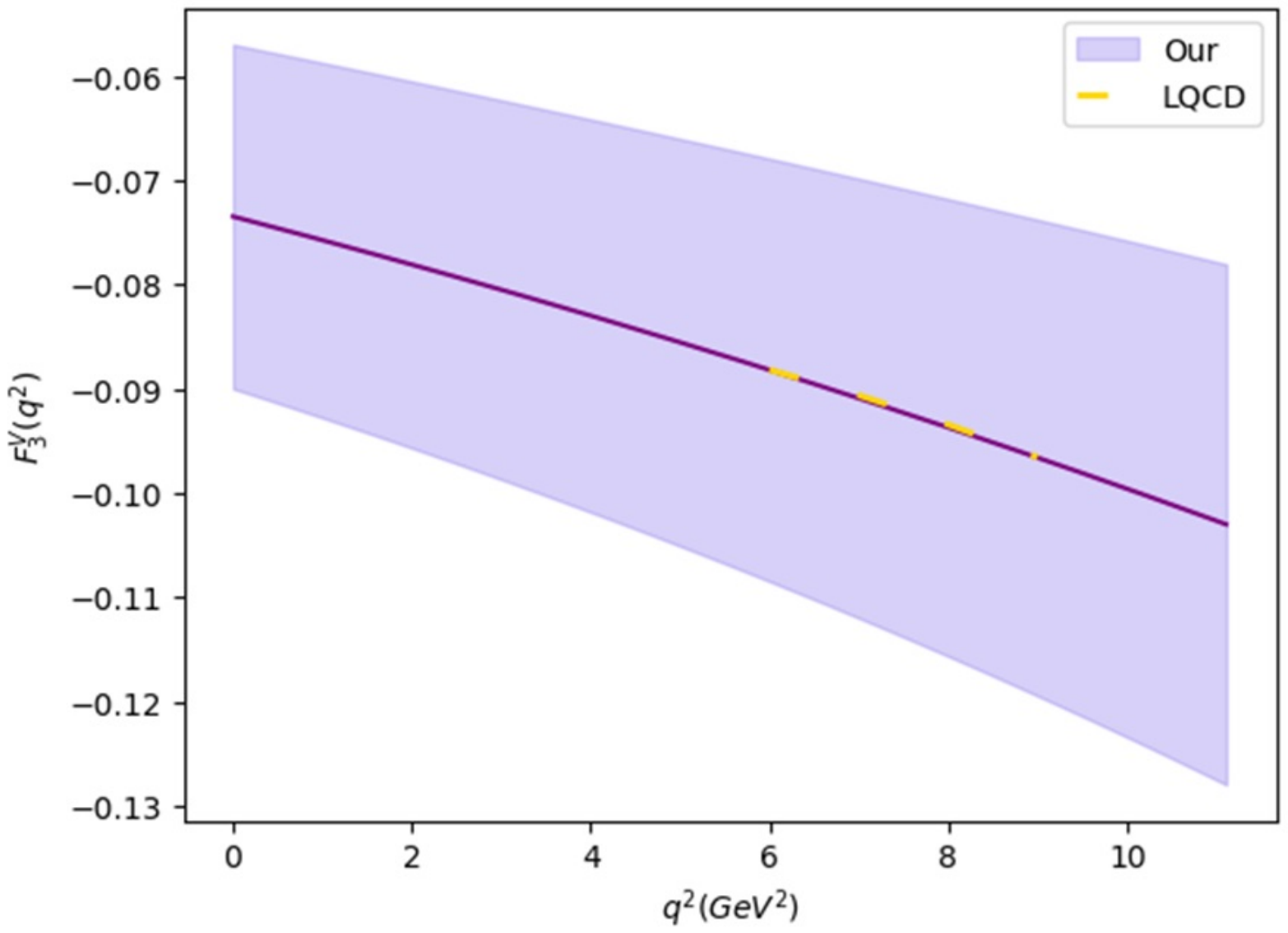}
     \end{subfigure}
     \hfill
          \begin{subfigure}[b]{0.49\textwidth}
         \centering
         \includegraphics[width=\textwidth]{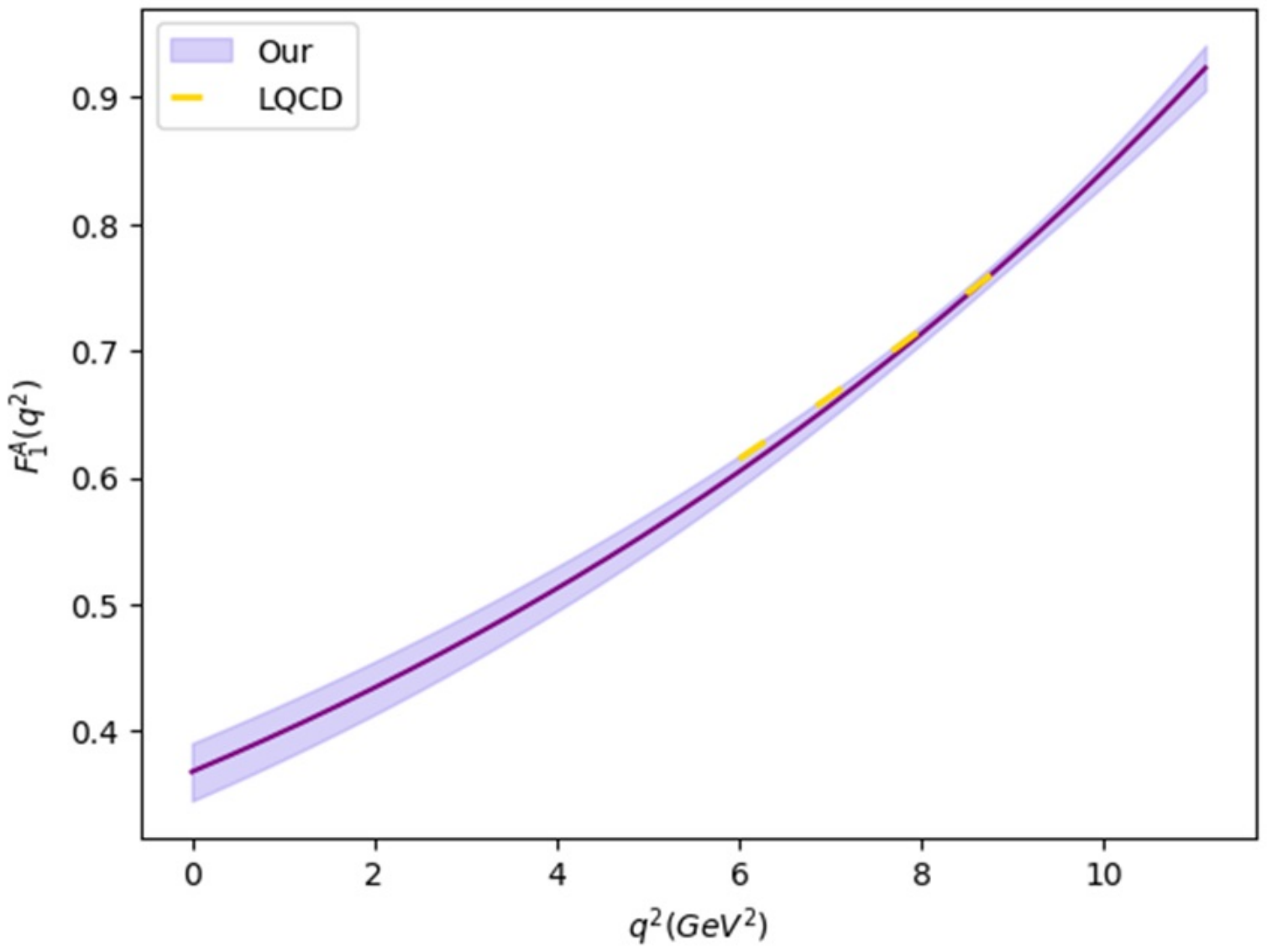}
     \end{subfigure}
     \hfill
              \begin{subfigure}[b]{0.49\textwidth}
         \centering
         \includegraphics[width=\textwidth]{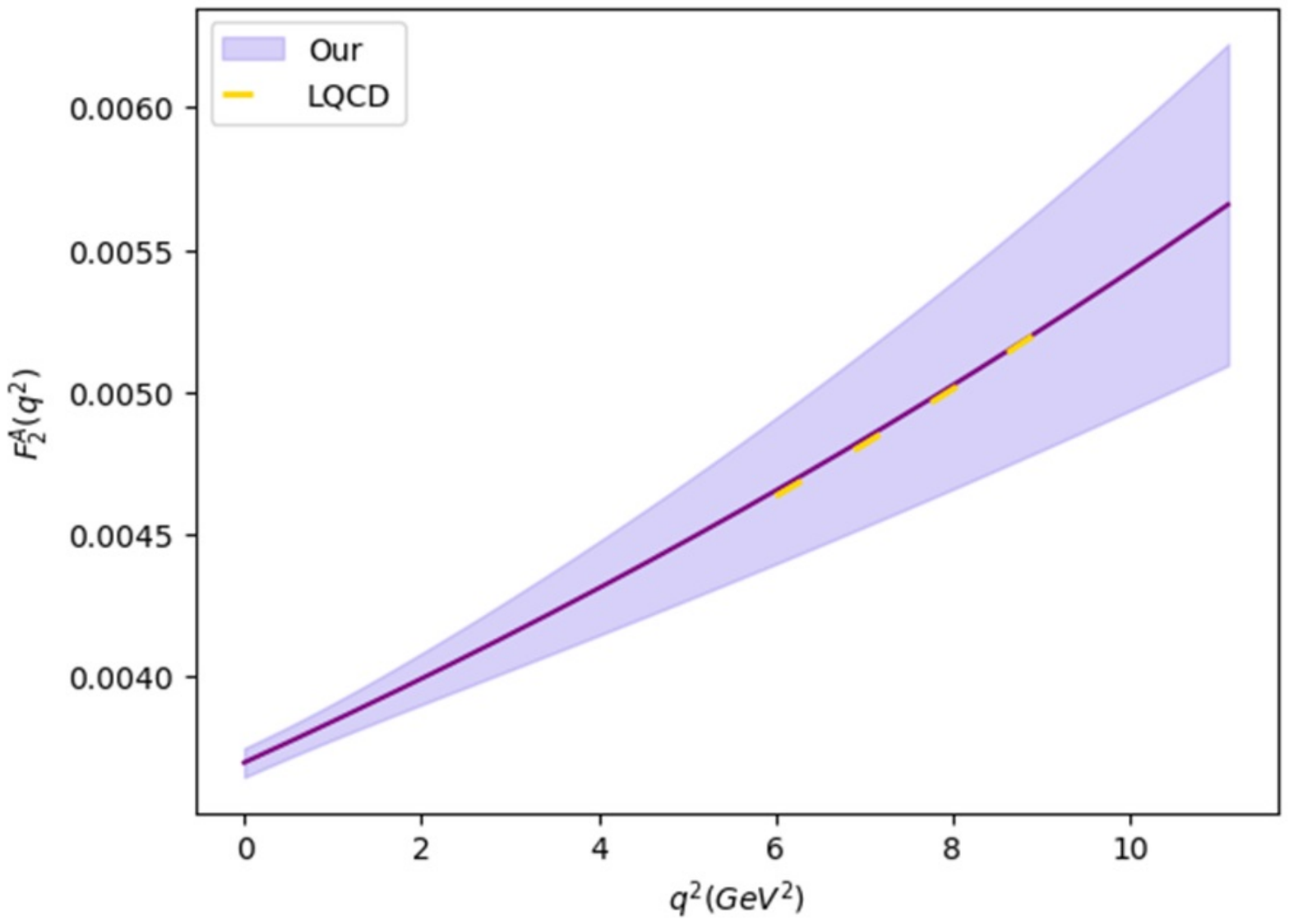}
     \end{subfigure}
     \hfill
              \begin{subfigure}[b]{0.49\textwidth}
         \centering
         \includegraphics[width=\textwidth]{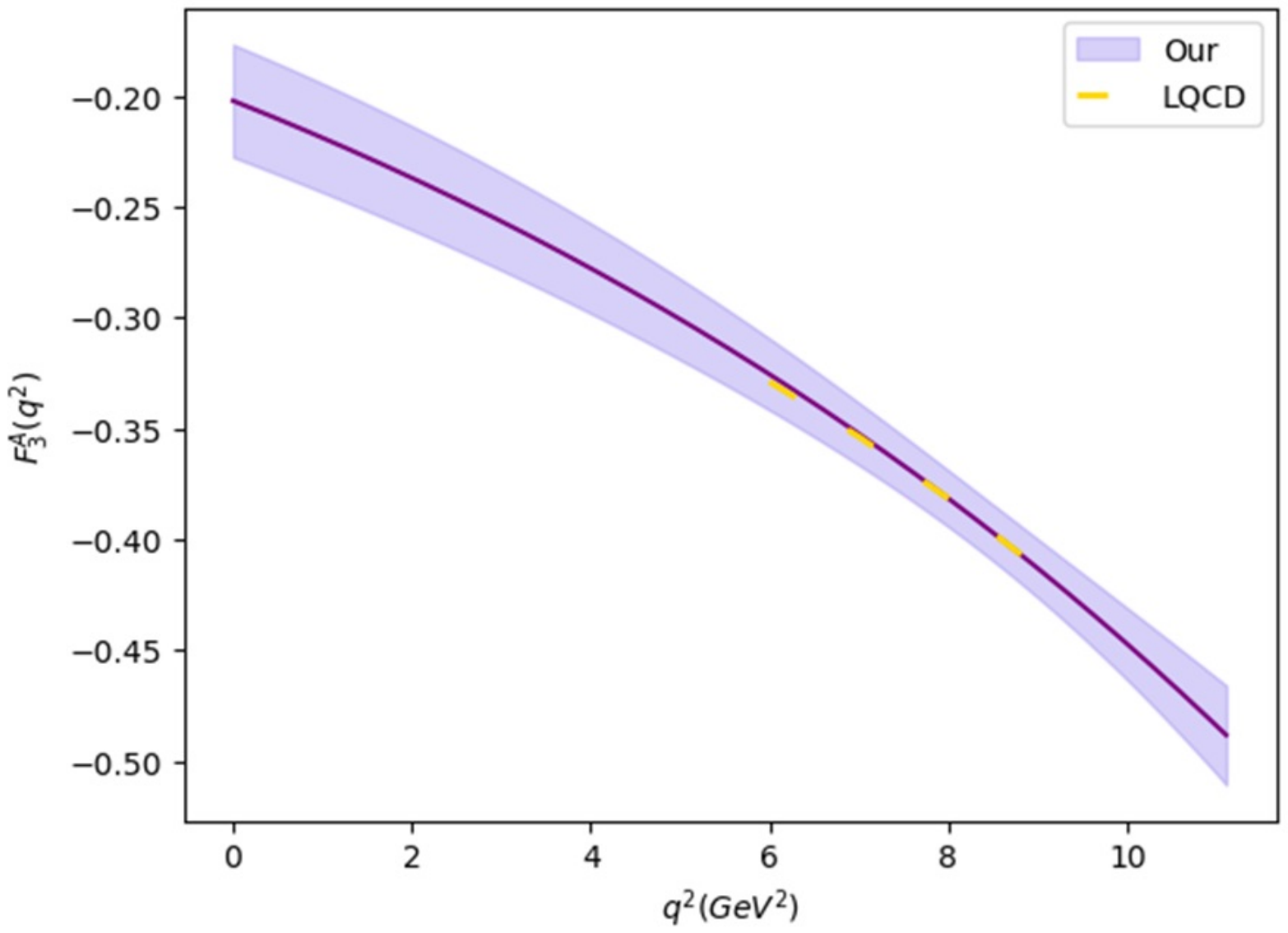}
     \end{subfigure}
     \hfill
      \caption{Phase-moment form factor fits to LQCD data for transition $\Lambda_b \to \Lambda_c$.} 
      \label{fig: fits}
\end{figure}

This section is devoted to the numerical calculations of the quantities obtained in the previous subsections. The input values for the masses of baryons and leptons are adopted from PDG as follows:
$M_{\Lambda_b} = 5.6196 GeV, M_{\Lambda_c} = 2.2865 GeV,$
$m_e = 0.511 MeV, m_\mu = 105.658 MeV, m_\tau = 1.777 GeV$ in addition to the CKM matrix element, the life time of $\Lambda_b^0$, and Fermi coupling constant parameters, $V_{cb} = 0.0408, {\tau _{{\Lambda _b}}} = 1.471ps$, and $G_F = 1.166 \times 10^{-5} GeV^{-2}$ \cite{ParticleDataGroup:2022pth}. The form factors defined in the LQCD formalism \cite{Detmold:2015aaa} had been expressed to those Weinberg functions $F_i^{(V,A)}$,
\begin{equation}
  {f_ + }({q^2}) = F_1^V({q^2}) + \frac{{{q^2}}}{{{M_1}({M_1} + {M_2})}}F_2^V({q^2}), \hfill \\
\end{equation}
\begin{equation}
  {f_ \bot }({q^2}) = F_1^V({q^2}) + \frac{{{M_1} + {M_2}}}{{{M_1}}}F_2^V({q^2}), \hfill \\
\end{equation}
\begin{equation}
  {f_0}({q^2}) = F_1^V({q^2}) + \frac{{{q^2}}}{{{M_1}({M_1} - {M_2})}}F_3^V({q^2}), \hfill \\
\end{equation}
\begin{equation}
  {g_ + }({q^2}) = F_1^A({q^2}) - \frac{{{q^2}}}{{{M_1}({M_1} - {M_2})}}F_2^A({q^2}), \hfill \\
\end{equation}
\begin{equation}
  {g_ \bot }({q^2}) = F_1^A({q^2}) - \frac{{{M_1} - {M_2}}}{{{M_1}}}F_2^A({q^2}), \hfill \\
\end{equation}
\begin{equation}
  {g_0}({q^2}) = F_1^A({q^2}) - \frac{{{q^2}}}{{{M_1}({M_1} + {M_2})}}F_3^A({q^2}). \hfill \\ 
\end{equation} 
With these equations of form factors and the LQCD data, we performed a fit to determine the free parameters in our formalism. Our results are shown in Figs. \ref{fig: fits}. Lattice group have done the extrapolation to the physical limit for the transition $\Lambda_b \to \Lambda_c$ by,
\begin{equation}
    f({q^2}) = \frac{1}{{1 - {q^2}/{{(m_{pole}^f)}^2}}}[a_0^f + a_1^f{z^f}({q^2})]
\end{equation}
where $m_{pole}^{{f_ + },{f_ \bot }} = 6.332,m_{pole}^{{f_0}} = 6.725,m_{pole}^{{g_ + },{g_ \bot }} = 6.768,m_{pole}^{{g_0}} = {\text{6}}{\text{.276}}$ (all in $GeV)$ \cite{Detmold:2015aaa}, and the central
values and uncertainties of the parameters from the nominal form factor parameters {$a_0^f, a_1^f$} are taken from Ref. \cite{Detmold:2015aaa}. The expansion parameter $z^f(q^2)$ is defined as
\begin{equation}
   {z^f}({q^2}) = \frac{{\sqrt {\mathcal{S}_ + ^f - {q^2}}  - \sqrt {\mathcal{S}_ + ^f - {\mathcal{S}_0}} }}{{\sqrt {\mathcal{S}_ + ^f - {q^2}}  + \sqrt {\mathcal{S}_ + ^f - {\mathcal{S}_0}} }},
\end{equation}
where one can choose ${\mathcal{S}_0} = {({M_1} - {M_2})^2}$, and $\mathcal{S}_ + ^f({\Lambda _b} \to {\Lambda _c}) = {(m_{pole}^f)^2}$. In Figs. \ref{fig: fits}, We determined the total average $\chi^2$ in our fits as $0.541$. Our fitted values for the parameters of the form factors are tabulated in Table \ref{Table: fit parameters-PM}. \par
Furthermore, Boyd-Grinstein-Lebed
parametrization (BGL) can be expressed \cite{Boyd:1997kz}
\begin{equation}
    F(t) = \frac{1}{{P(t)\phi ({z'})}}\sum\limits_{n = 0}^\infty  {{a_n}Z{{(t,{t_0})}^n}} 
\end{equation}
in the semileptonic region, where $Z(t,{t_s}) = \frac{{{t_s} - t}}{{{{(\sqrt {{t_ + } - t}  + \sqrt {{t_ + } - {t_s}} )}^2}}}$ and $t \equiv {q^2}$. The Blaschke factor $P(t)$ is defined by the masses of $B_c$ resonances below the $\Lambda_b\Lambda_c$ pair-production threshold. It can be written as \cite{Boyd:1997kz}
\begin{equation}
    P(t) = \prod\limits_p^{} {Z(t,{t_p})} 
\end{equation}
with $t_p$ evaluated at the invariant mass squared of each resonance. $B_c$ pole masses are taken from Ref. \cite{Boyd:1997kz}. The weighting function $\phi (z')$ can be assumed as following \cite{Boyd:1994tt}
\begin{equation}
    \phi (z') = \frac{{{2^{5/2}}{m_b}}}{{\sqrt 3 }}\frac{{{{({t_ + } - {t_ - })}^{ - 1/2}}{{(1 + z')}^2}}}{{{{(1 - z')}^{9/2}}}}{\left( {\beta  + \frac{{1 + z'}}{{1 - z'}}} \right)^5},
\end{equation}
where $\beta  = \sqrt {{t_ + }/({t_ + } - {t_ - })} $, $m_b = 4.9 GeV$, and $\sqrt {\frac{{{t_ + } - {q^2}}}{{{t_ + } - {t_ - }}}}  = \frac{{1 + z'}}{{1 - z'}}$. We have reported our values for the first two free parameters of the BGL formula in Table \ref{Table: fit parameters-BGL} by fitting to the LQCD formalism \cite{Detmold:2015aaa}. The total average $\chi^2$ is $0.504$. 
\par
\begin{figure}
     \centering
     \begin{subfigure}[b]{0.49\textwidth}
         \centering
         \includegraphics[width=\textwidth]{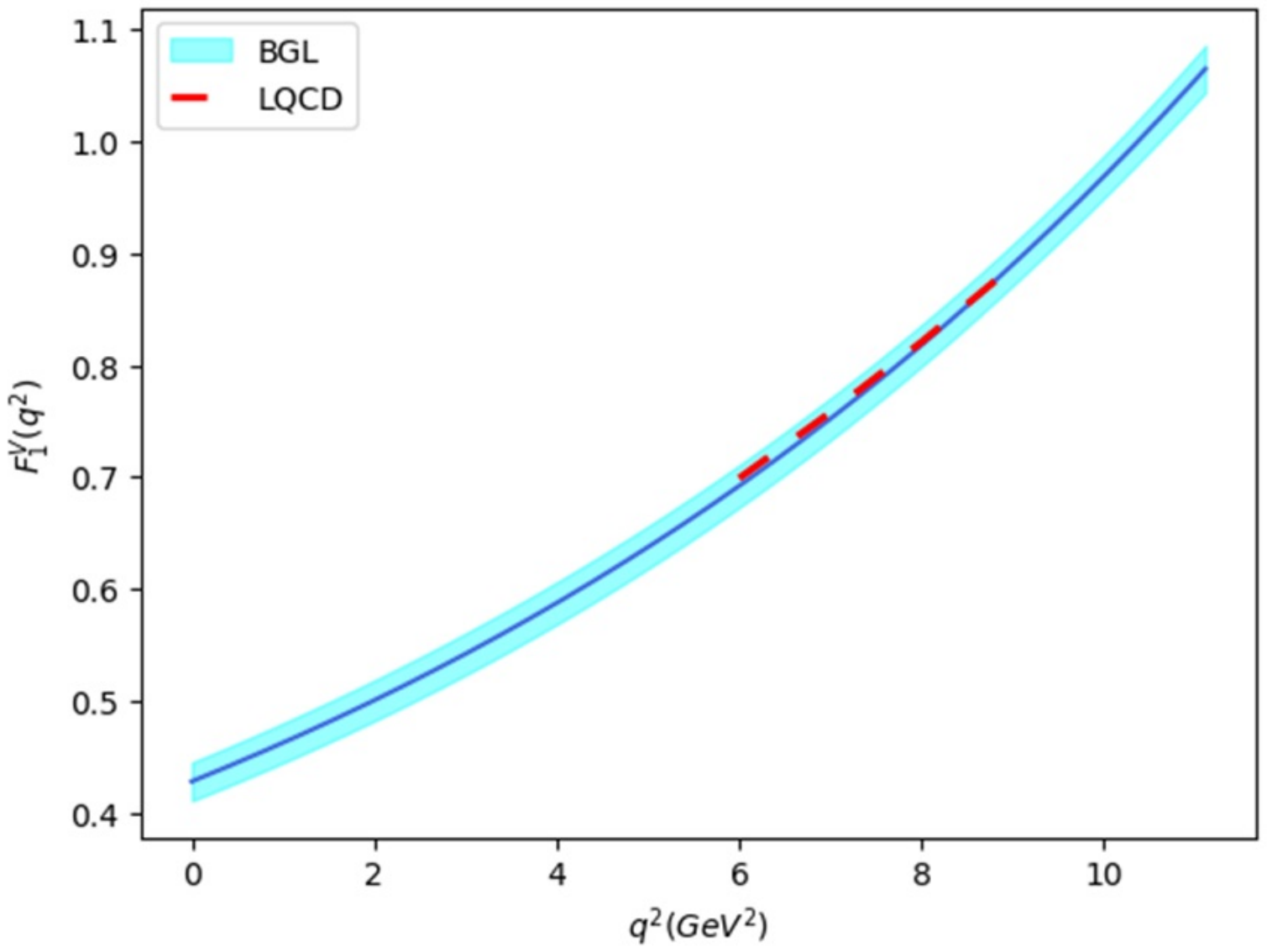}
     \end{subfigure}
     \hfill
     \begin{subfigure}[b]{0.49\textwidth}
         \centering
         \includegraphics[width=\textwidth]{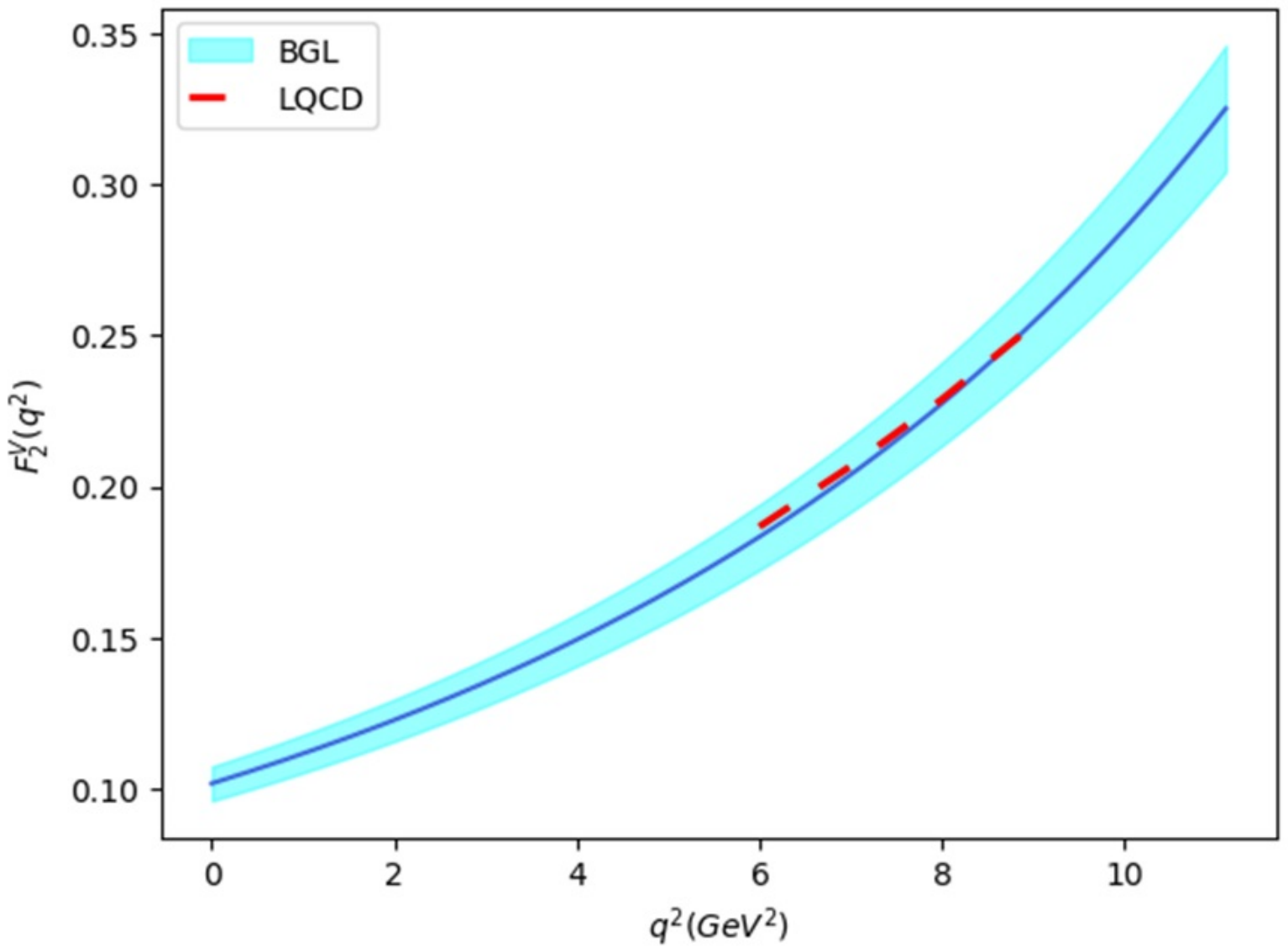}
     \end{subfigure}
     \hfill
     \begin{subfigure}[b]{0.49\textwidth}
         \centering
         \includegraphics[width=\textwidth]{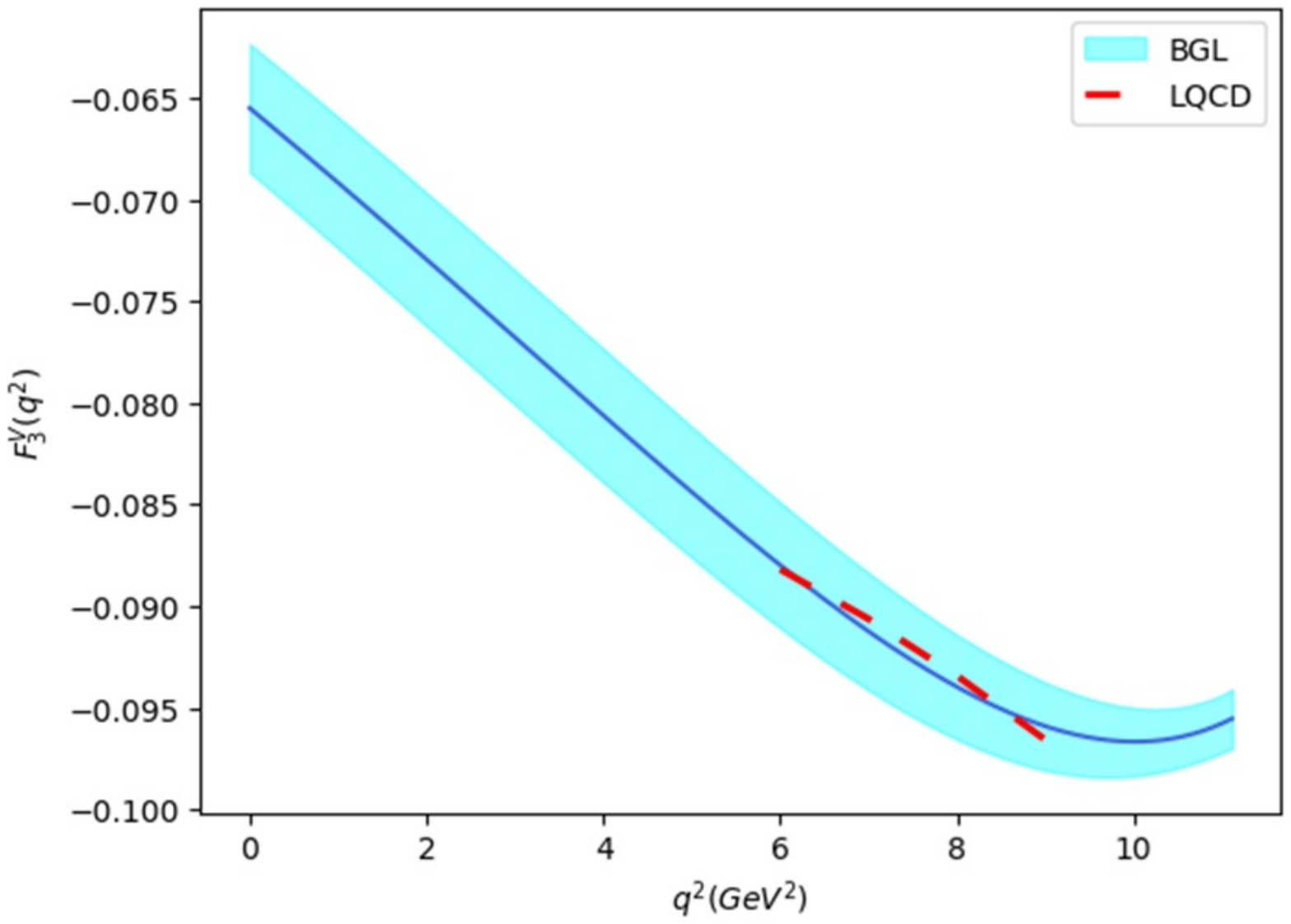}
     \end{subfigure}
     \hfill
          \begin{subfigure}[b]{0.49\textwidth}
         \centering
         \includegraphics[width=\textwidth]{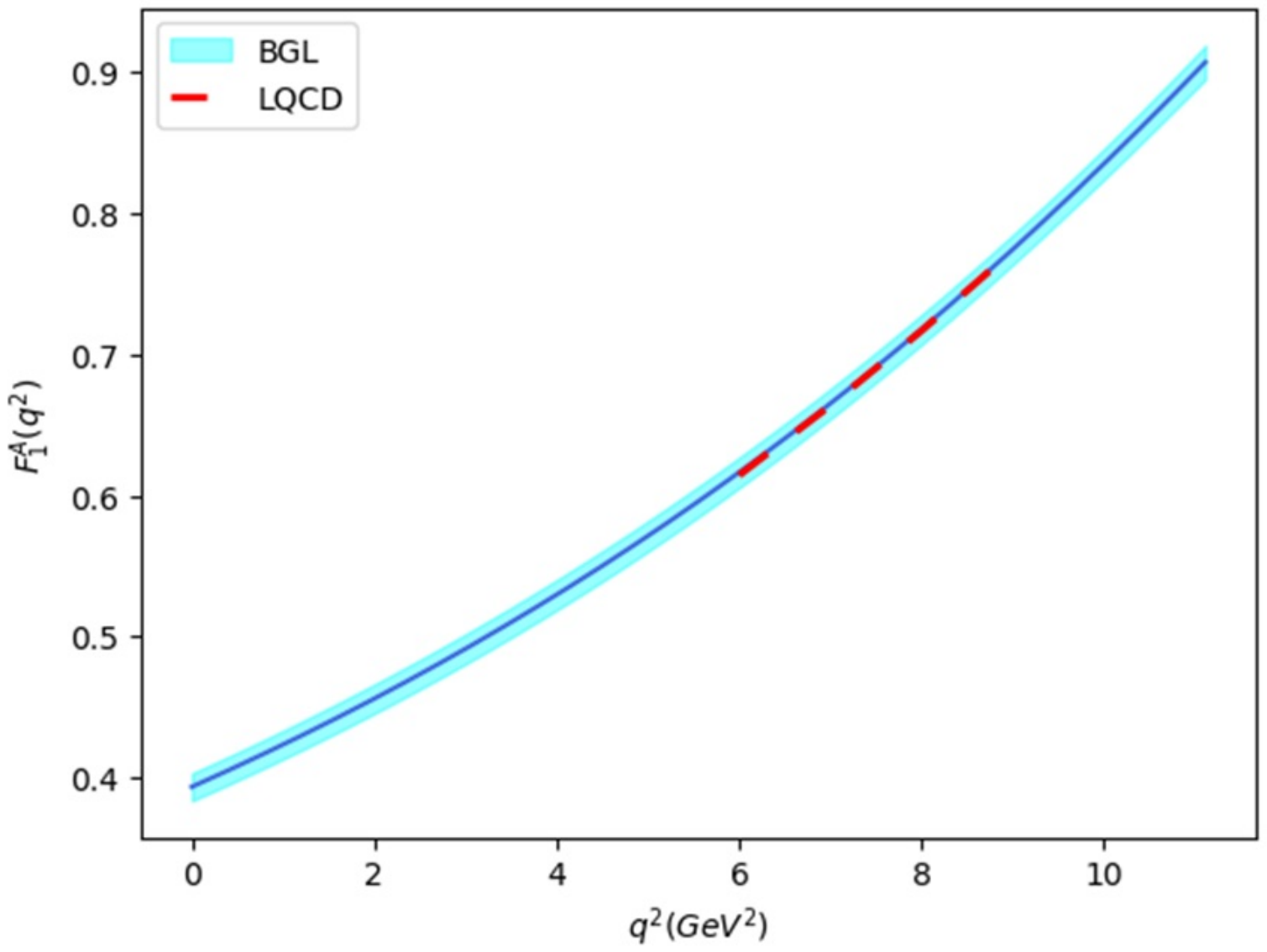}
     \end{subfigure}
     \hfill
              \begin{subfigure}[b]{0.49\textwidth}
         \centering
         \includegraphics[width=\textwidth]{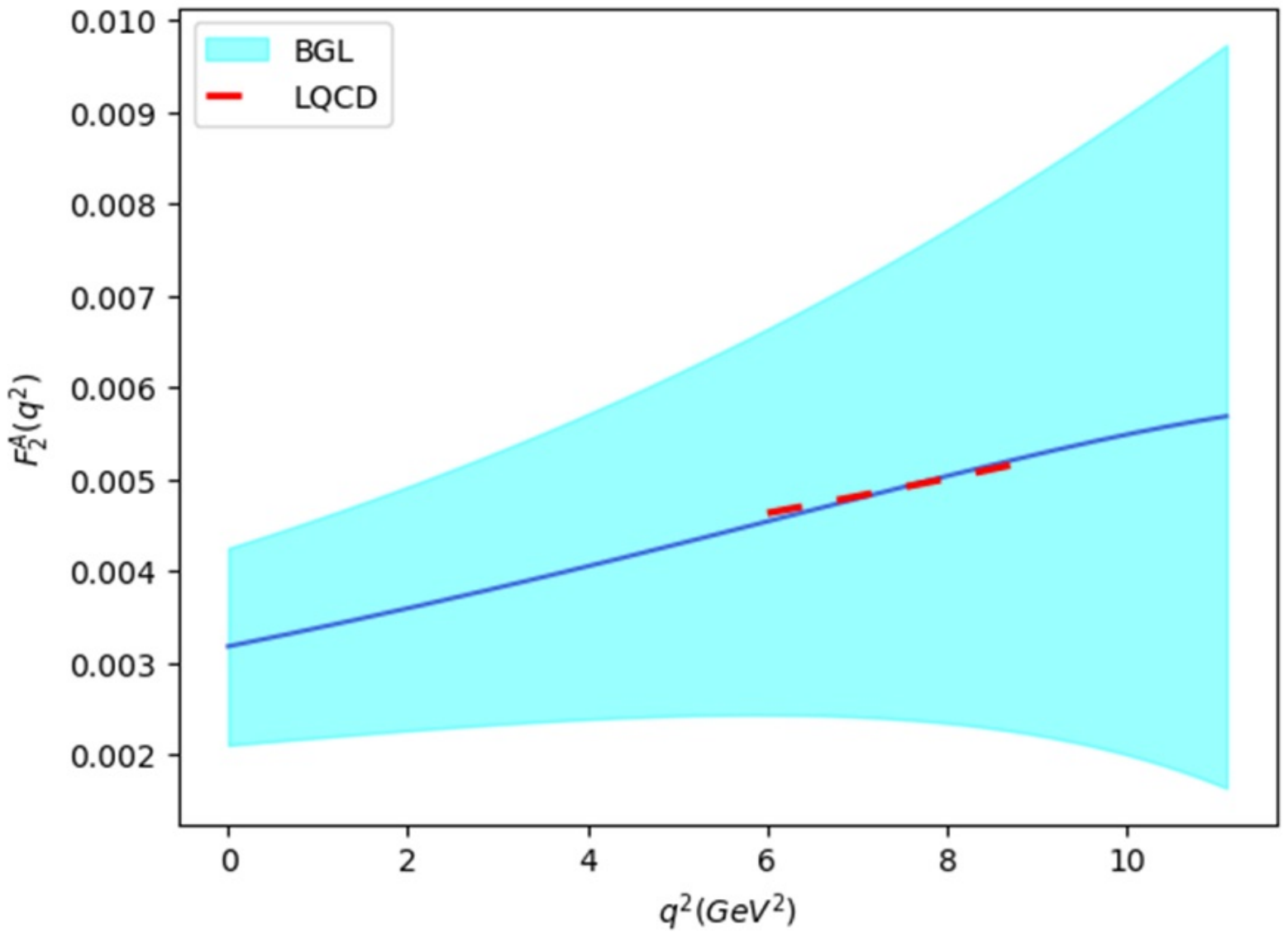}
     \end{subfigure}
     \hfill
              \begin{subfigure}[b]{0.49\textwidth}
         \centering
         \includegraphics[width=\textwidth]{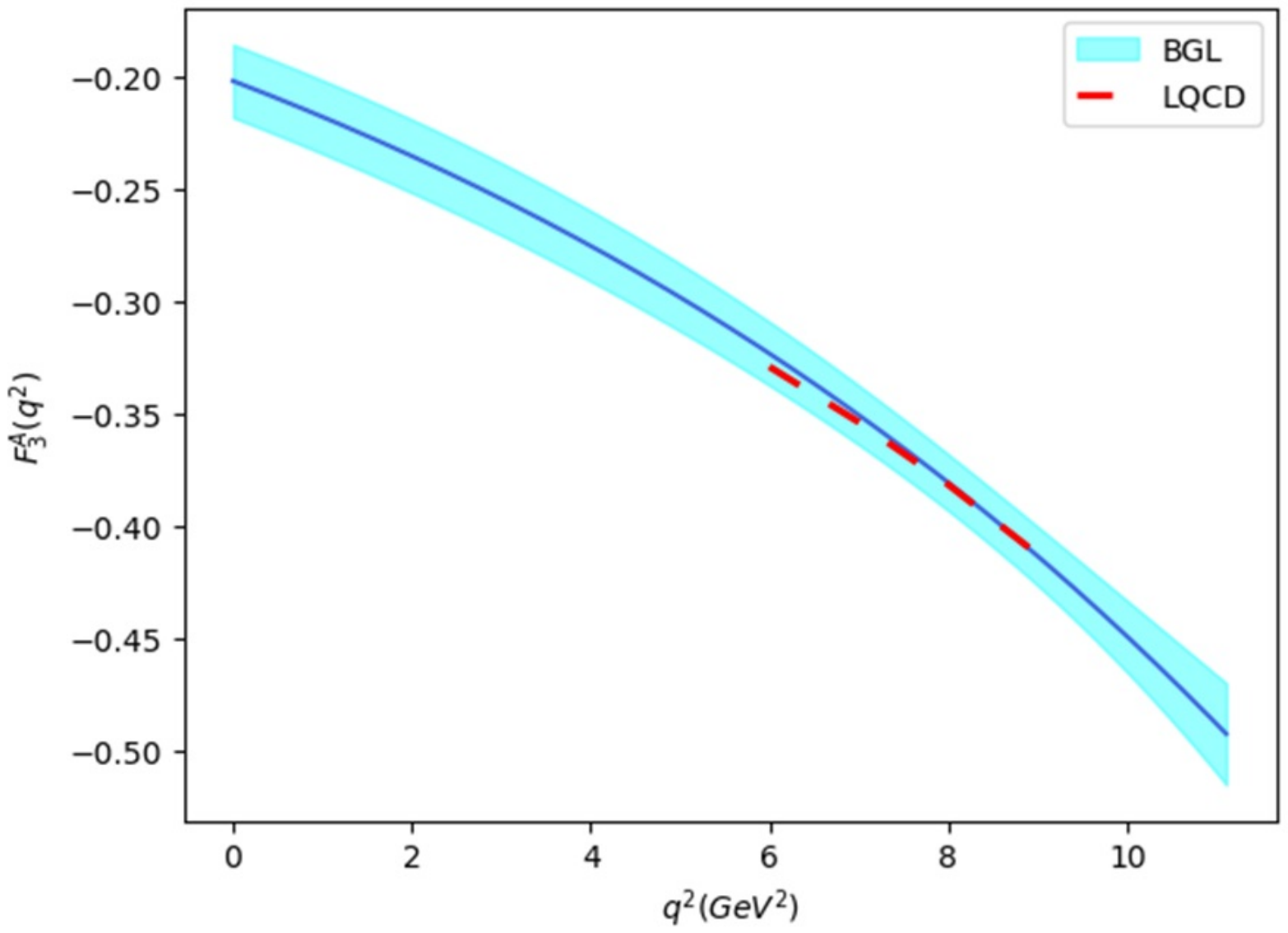}
     \end{subfigure}
     \hfill
      \caption{BGL form factor fits to LQCD data for transition $\Lambda_b \to \Lambda_c$.} 
      \label{fig: fits-BGL}
\end{figure}

In Figs. \ref{fig: fits-BGL}, we have plotted our results for fits of the BGL form factor parameterization to LQCD. As can be seen, in some cases such as $F_3^V(q^2)$ and $F_3^A(q^2)$ the behavior of BGL form factors can not describe the LQCD data well. Also by increasing $q^2$, the figure of $F_3^V(q^2)$ will be strange, see $q^2 > 8 GeV^2$. PM parameterization is consistent with the expectations of form factor patterns. $F_1^V, F_2^V, F_1^A, F_2^A$ gradually increase by growing $q^2$, and $F_3^{V,A}$ have opposite tendencies while $q^2$ enhance, see Figs. \ref{fig: fits}.
\par
The figures of PM and BGL form factors are almost similar near $q^2=0$. Our values in Table \ref{Table: q2=0} are related to the form factors at $q^2 = 0$ with both PM and BGL models. They are compatible with those obtained in Ref. \cite{Gutsche:2015mxa} in which the authors reported $F_1^V(0) = 0.549, F_2^V(0) = 0.110, F_3^V(0) =  - 0.023, F_1^A(0) = 0.542, F_2^A(0) = 0.018, F_3^A(0) =  - 0.123$ with the
double-pole parametrization. Zhu {\it{et} \it{al.}} resulted the following central values of $\Lambda_b \to \Lambda_c$ transition form factors $F_1^V(0) = 0.500,F_2^V(0) = 0.098,F_3^V(0) =  - 0.009,F_1^A(0) = 0.509,F_2^A(0) = 0.015,F_3^A(0) =  - 0.085$ in their covariant light-front approach \cite{Zhu:2018jet}.

\begin{table}[htb]
\caption{Parameters of the PM form factors for the $\Lambda_b \to \Lambda_c$ transitions.}
\label{Table: fit parameters-PM}
\begin{tabular}{c|c|c|c|c|c|c}
\hline\hline

 & $F_1^V$ & $F_2^V$ & $F_3^V$ & $F_1^A$ & $F_2^A$ & $F_3^A$
\\
\hline
${\mathcal{F}(0)}$ & $0.429 \pm 0.026$ & $0.101 \pm 0.025$ & $ - 0.073 \pm 0.016$ & $0.368 \pm 0.023$ & $0.004 \pm 0.00$ & $ - 0.202 \pm 0.025$
\\
\hline
$\mathcal{A}_0^{\mathcal{F}}(0)$ & $5.061 \pm 0.407$ & $6.361 \pm 1.635$ & $1.9 \pm 0.512$ & $5.171 \pm 0.437$ & $2.390 \pm 0.555$ & $4.958 \pm 0.868$
\\
\hline\hline
\end{tabular}
\end{table}

\begin{table}[htb]
\caption{Parameters of the BGL form factors for the $\Lambda_b \to \Lambda_c$ transitions.}
\label{Table: fit parameters-BGL}
\begin{tabularx}{1\textwidth}{>{\centering\arraybackslash}X | >{\centering\arraybackslash}X | >{\centering\arraybackslash}X | >{\centering\arraybackslash}X | >{\centering\arraybackslash}X | >{\centering\arraybackslash}X | >{\centering\arraybackslash}X }
\hline\hline

 & $F_1^V$ & $F_2^V$ & $F_3^V$ & $F_1^A$ & $F_2^A$ & $F_3^A$
\\
\hline
$a_0$ & \makecell{0.0798 \\ $\pm$ 0.0016} & \makecell{0.0244 \\ $\pm$ 0.0016} & \makecell{$-0.0072$ \\ $\pm$ 0.0001} & \makecell{0.0702 \\ $\pm$ 0.0009} & \makecell{0.0004 \\ $\pm$ 0.0003} & \makecell{$-0.0382$ \\ $\pm$ 0.0018}
\\
\hline
$a_1$ & \makecell{0.90 \\ $\pm$ 0.0955} & \makecell{0.1000 \\ $\pm$ 0.0098}  & \makecell{$-0.2400$ \\ $\pm$ 0.0185} & \makecell{0.9171 \\ $\pm$ 0.0546} & \makecell{0.0010 \\ $\pm$ 0.0000} & \makecell{$-0.4228$ \\ $\pm$ 0.1162} 
\\
\hline\hline
\end{tabularx}
\end{table}

\begin{table}[htb]
\caption{Form factors values at $q^2 = 0$.}
\label{Table: q2=0}
\begin{tabularx}{1\textwidth}{>{\centering\arraybackslash}X | >{\centering\arraybackslash}X | >{\centering\arraybackslash}X | >{\centering\arraybackslash}X | >{\centering\arraybackslash}X | >{\centering\arraybackslash}X | >{\centering\arraybackslash}X }
\hline\hline

 & $F_1^V$ & $F_2^V$ & $F_3^V$ & $F_1^A$ & $F_2^A$ & $F_3^A$
\\
\hline
${\mathcal{F}(0)}$ (PM) & $0.429 \pm 0.026$ & $0.101 \pm 0.025$ & $ - 0.073 \pm 0.016$ & $0.368 \pm 0.023$ & $0.004 \pm 0.00$ & $ - 0.202 \pm 0.025$
\\
\hline
${\mathcal{F}(0)}$ (BGL) & $0.429 \pm 0.017$ & $0.101 \pm 0.006$ & $-0.065 \pm 0.003$ & $0.394 \pm 0.010$ & $0.003 \pm 0.001$ & $-0.201 \pm 0.016$
\\
\hline\hline
\end{tabularx}
\end{table}

\begin{figure}[h]
    \includegraphics[width=0.5\textwidth]{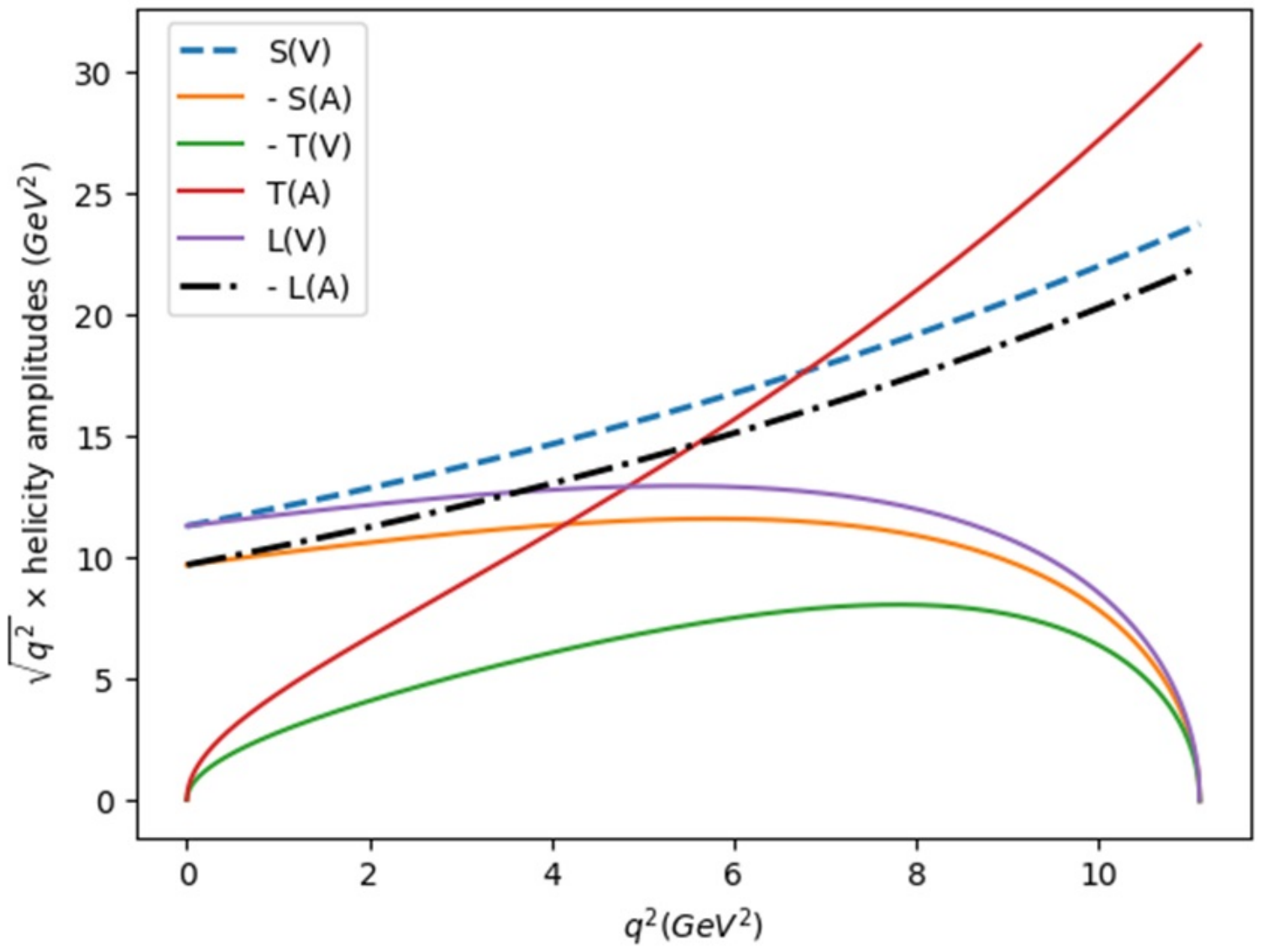}
    \caption{The six independent helicity amplitudes $S(V,A) = \sqrt {{q^2}} H_{\frac{1}{2},t}^{V,A},T(V,A) = \sqrt {{q^2}} H_{\frac{1}{2},1}^{V,A},$ and $L(V,A) = \sqrt {{q^2}} H_{\frac{1}{2},0}^{V,A}$ multiplied by $\sqrt{q^2}$.}
    \label{fig: helicity}
\end{figure}

The dominant form factors are $F_1^V$ and $F_1^A$ with bigger magnitudes than other ones, and their $q^2$ dependencies are close for both PM and BGL configurations. If we consider these form factors at the zero recoil point, where $q^2 = q^2_{max}$, one can calculate $F_1^V(q^2_{max}) = 1.054 \pm 0.024$, $F_1^A(q^2_{max}) = 0.923 \pm 0.018$ with PM, and $F_1^V(q^2_{max}) = 1.064 \pm 0.021$, and $F_1^A(q^2_{max}) = 0.906 \pm 0.012$ by BGL, which are close to 1, which would satisfy the prediction of HQET. Moreover, they are near the values of 0.985 and 0.966 in Ref. \cite{Gutsche:2015mxa}. In Fig. \ref{fig: helicity}, we represent the $q^2$ dependence of the six helicity amplitudes based on the considered invariant form factors with the PM model. The longitudinal and scalar helicity amplitudes dominate near $q^2 = 0 $, where $H_{\frac{1}{2},0}^{V,A} \approx H_{\frac{1}{2},t}^{V,A}$. Close to the zero recoil point $q^2 = (M_{\Lambda_b} -M_{\Lambda_c})^2$, the orbital S-wave contributions of $H_{\frac{1}{2},1}^A, H_{\frac{1}{2},t}^V$ are the dominant amplitudes.

\par
\begin{figure}
     \centering
     \begin{subfigure}[b]{0.49\textwidth}
         \centering
         \includegraphics[width=\textwidth]{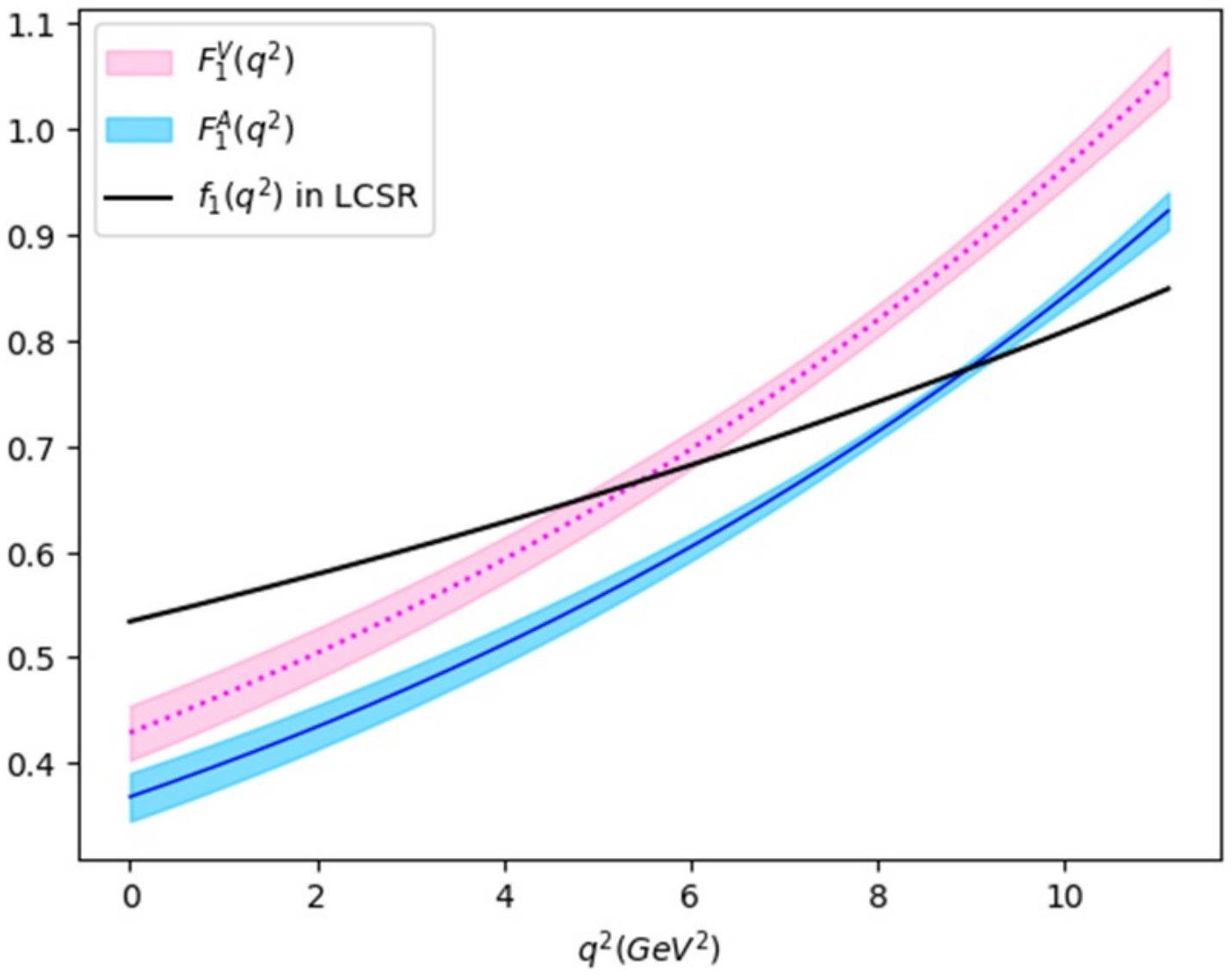}
     \end{subfigure}
     \hfill
     \begin{subfigure}[b]{0.49\textwidth}
         \centering
         \includegraphics[width=\textwidth]{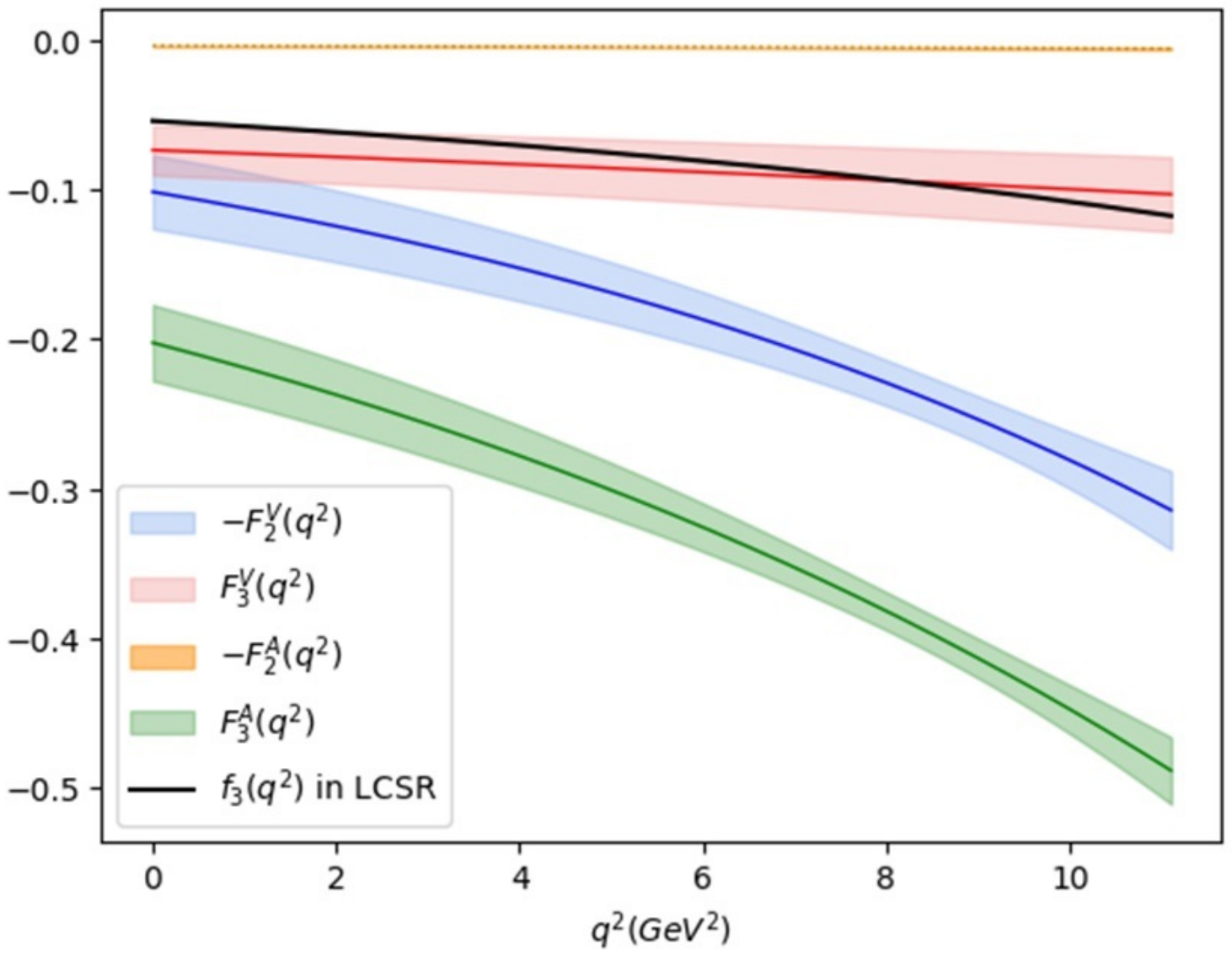}
        
     \end{subfigure}
     \caption{The form factors $F_1^V(q^2)$, $F_1^A(q^2)$, and $F_2^{(V,A)}(q^2)$, $F_3^{(V,A)}(q^2)$ in the phase-moment parameterization compared with LCSR.}
      \label{fig: F2_compare_LCSR}
     \hfill
     
\end{figure}

To compare our results with the LCSR \cite{Duan:2022uzm} approach, we plotted our form factors with their error band in the PM scheme and those obtained in LCSR in Figs \ref{fig: F2_compare_LCSR}. Note that their form factors are identified as $f_1(q^2), f_2(q^2), f_3(q^2), g_1(q^2), g_2(q^2), g_3(q^2),$ in which there are the following relations, $f_1(q^2) = g_1(q^2)$, and $f_2(q^2) = f_3(q^2) = g_2(q^2) = g_3(q^2)$ with the LCSR procedure. Then, they got the form factors using the $Z$-expansion formula. Our defined form factors in the section \ref{section: helicity} are analogous to them as $F_1^V = f_1, F_2^V = -f_2, F_3^V = f_3, F_1^A = g_1, F_2^A = -g_2,$ and $F_3^A = g_3$. The black lines show $f_1$ and $f_3$ in LCSR. Our figures are different from theirs. In their LCSR calculation procedure with the quark-hadron duality assumption, they have taken only two form factors; however, the contribution of other form factors is important in acquiring the physical observables such as decay rates.

Substituting the fitting parameters of the form factors into the helicity form of decay widths, then integrating over the momentum square $q^2$ on the whole physical region, the information on the decay widths can be evaluated. We report the ratios and decay rates in Tables \ref{Table: BR} to \ref{Table: partial decay widths hf}. The results for $\mu$ and $e$ lepton channels are close but differ from $\tau$ mode. 

\begin{table}[htb]

\caption{Decay widths in $10^{(-15)} GeV$ and branching ratios in percentage for semileptonic transition of $\Lambda_b \to \Lambda_c$.}
\label{Table: BR}
\begin{tabularx}{1\textwidth}{>{\centering\arraybackslash}X | >{\centering\arraybackslash}X | >{\centering\arraybackslash}X |>{\centering\arraybackslash}X |>{\centering\arraybackslash}X |>{\centering\arraybackslash}X }

\hline\hline

Decay mode & Decay width (PM) & Decay width (BGL) & BR(PM) & BR(BGL) & BR(PDG) \cite{ParticleDataGroup:2022pth}
\\
\hline
$\Lambda _b^0 \to \Lambda _c^ + {e^ - }{\bar \nu _e}$ & $23.306 \pm 0.773$ & $23.798 \pm 0.736$& $5.211$ & $5.321$ & $ (6.2_{ - 1.3}^{ + 1.4})\% $ 
\\
\hline
$\Lambda _b^0 \to \Lambda _c^ + {\mu^ - }{\bar \nu _\mu}$ & $23.236\pm 1.999$ & $23.724 \pm 1.960$ & $5.195$ & $5.305$ &
\\
\hline
$\Lambda _b^0 \to \Lambda _c^ + {\tau^ - }{\bar \nu _\tau}$ & $7.798 \pm 0.238$ & $7.836 \pm 0.229 $ & $1.744$ & $1.752$ &
\\
\hline\hline
\end{tabularx}
\end{table}

\begin{table}[htb]
\caption{$q^2$ averages of helicity structure functions for nonflip contributions ($10^{(-15)} GeV$).}
\label{Table: partial decay widths nf}
\begin{tabularx}{1\textwidth}{>{\centering\arraybackslash}X | >{\centering\arraybackslash}X | >{\centering\arraybackslash}X | >{\centering\arraybackslash}X | >{\centering\arraybackslash}X | >{\centering\arraybackslash}X | >{\centering\arraybackslash}X }

\hline\hline

Mode & $\Gamma_U$ & $\Gamma_L$ & $\Gamma_{LT}$ & $\Gamma_{P}$ & $\Gamma_{L_P}$ & $\Gamma_{{LT}_P}$
\\
\hline
$e$ (PM) & 9.48 & 13.83 & 4.93   & 6.41 & 13.15 & -2.54 
\\
\hline
$e$ (BGL) & 9.65 & 14.15 & 5.21   & 6.48 & 13.48 & -2.59 
\\
\hline
$\mu$ (PM) & 9.44 & 13.63 & 4.91 & 6.38 & 12.96	& -2.52
\\
\hline
$\mu$ (BGL) & 9.61 & 13.94 & 5.19 & 6.45 & 13.27	& -2.57
\\
\hline
$\tau$ (PM) & 2.71 &	2.39 &	1.53	& 1.61 &	2.11 &	-0.54
\\
\hline
$\tau$ (BGL) & 2.72 &	2.40 &	1.54	& 1.62 &	2.12 &	-0.54
\\
\hline\hline
\end{tabularx}
\end{table}

\begin{table}[htb]
\caption{$q^2$ averages of helicity structure functions considering helicity flip contributions (in $10^{(-15)} GeV$). We do not display helicity flip results for the $e$ mode, because they are too small in the above units.}
\label{Table: partial decay widths hf}
\begin{tabularx}{1\textwidth}{>{\centering\arraybackslash}X | >{\centering\arraybackslash}X | >{\centering\arraybackslash}X | >{\centering\arraybackslash}X | >{\centering\arraybackslash}X | >{\centering\arraybackslash}X | >{\centering\arraybackslash}X | >{\centering\arraybackslash}X |>{\centering\arraybackslash}X |>{\centering\arraybackslash}X |>{\centering\arraybackslash}X |>{\centering\arraybackslash}X }

\hline\hline

Mode & ${\tilde \Gamma _U}$ & ${\tilde \Gamma _L}$ & ${\tilde \Gamma _S}$ & ${\tilde \Gamma _{LT}}$  & ${\tilde \Gamma _{S_P}}$ & ${\tilde \Gamma _{SL}}$ & ${\tilde \Gamma _{P}}$ & ${\tilde \Gamma _{L_P}}$ & ${\tilde \Gamma _{{LT}_P}}$ & ${\tilde \Gamma _{{ST}_P}}$ & ${\tilde \Gamma _{{SL}_P}}$ 
\\
\hline
$\mu$ (PM) & 0.01 & 0.04	& 0.04 &	0.005	& 0.04 & 0.04 & 0.007 &	0.04 &	-0.004 & -0.007 & 0.04
\\
\hline
$\mu$ (BGL) & 0.01 & 0.04	& 0.04 &	0.006	& 0.04 & 0.04 & 0.007 &	0.04 &	-0.004 & -0.007 & 0.04
\\
\hline
$\tau$ (PM) & 0.54 &	0.51 &	0.55 &	0.29
& 0.47	& 0.46	& 0.33	& 0.46 &	-0.11 &	-0.34 & 0.53
\\
\hline
$\tau$ (BGL) & 0.54 &	0.51 &	0.55 &	0.30
& 0.46	& 0.47	& 0.34	& 0.47 &	-0.12 &	-0.34 & 0.53
\\
\hline\hline
\end{tabularx}
\end{table}

Our total decay widths, $\Gamma(\Lambda_b^0 \to \Lambda_c^+ e^- \bar {\nu}_e) = (23.306\pm 0.773) \times 10^{-15} GeV$ and $\Gamma(\Lambda_b^0 \to \Lambda_c^+ e^- \bar {\nu}_e) = (23.798\pm 0.736) \times 10^{-15} GeV$ are close to the covariant confined quark model in Ref. \cite{Gutsche:2015mxa} with $\Gamma(\Lambda_b^0 \to \Lambda_c^+ e^- \bar {\nu}_e) = 32.0 \times 10^{-15} GeV$. Our values for the total decay width of semi-electronic, semi-muonic and semi-tauonic satisfy the corresponding results $\Gamma (\Lambda _b^0 \to \Lambda _c^ + {e^ - }{{\bar \nu }_e}) = 2.60_{ - 0.54}^{ + 0.52},\Gamma (\Lambda _b^0 \to \Lambda _c^ + {\mu ^ - }{{\bar \nu }_\mu }) = 2.59_{ - 0.54}^{ + 0.52},\Gamma (\Lambda _b^0 \to \Lambda _c^ + {\tau ^ - }{{\bar \nu }_\tau }) = 0.71_{ - 0.13}^{ + 0.13},( \times {10^{ - 14}}GeV)$ by the LCSR method \cite{Duan:2022uzm}. According to our obtained quantities, one finds the branching fraction of
\begin{equation}
    \mathcal{R}(\Lambda _c^ + ) \equiv \frac{BR(\Lambda _b^0 \to \Lambda _c^ + {\tau ^ - }{\bar \nu _\tau })}{BR(\Lambda _b^0 \to \Lambda _c^ + {\mu ^ - }{\bar \nu _\mu })} = 0.336 \pm 0.031,
\end{equation}
with PM parameterization, and
\begin{equation}
    \mathcal{R}(\Lambda _c^ + ) \equiv \frac{BR(\Lambda _b^0 \to \Lambda _c^ + {\tau ^ - }{\bar \nu _\tau })}{BR(\Lambda _b^0 \to \Lambda _c^ + {\mu ^ - }{\bar \nu _\mu })} = 0.330 \pm 0.029,
\end{equation}
with BGL one, which are close to the values 0.333 and 0.33 reported by Detmold {\it{et al.}} \cite{Detmold:2015aaa} and Mu {\it{et al.}} \cite{Mu:2019bin} respectively. This quantity has been reported as $BR(\Lambda _b^0 \to \Lambda _c^ + {\tau ^ - }{\bar \nu _\tau })/BR(\Lambda _b^0 \to \Lambda _c^ + {\mu ^ - }{\bar \nu _\mu }) = 0.242 \pm 0.026 \pm 0.040 \pm 0.059$ by LHCb recently \cite{LHCb:2022piu}. The ratio $\mathcal{R}(\Lambda_c^+)$ obtained by our and other models are collected in Table \ref{Table: R}.

\begin{table}[htb]

\caption{$\mathcal{R}(\Lambda_c^+)$ compared with other approaches with the central values.}
\label{Table: R}
\begin{tabularx}{1\textwidth}{>{\centering\arraybackslash}X | >{\centering\arraybackslash}X | >{\centering\arraybackslash}X |>
{\centering\arraybackslash}X |>{\centering\arraybackslash}X |>{\centering\arraybackslash}X |>{\centering\arraybackslash}X |>{\centering\arraybackslash}X |>{\centering\arraybackslash}X}

\hline\hline

Models & This work (PM) & This work (BGL) & HQET \cite{Bernlochner:2018kxh, Bernlochner:2018bfn} & LQCD  \cite{Detmold:2015aaa} & Helicity basis \cite{Penalva:2020xup} & Light-front quark \cite{Li:2021qod} & LCSR \cite{Duan:2022uzm}& Covariant confined quark model \cite{Gutsche:2018nks}
\\
\hline
$\mathcal{R}(\Lambda_c^+)$ & $0.336 \pm 0.031$ & $0.330 \pm 0.029$ & 0.324 & 0.333 & 0.33 & 0.30 & 0.274 & 0.30
\\
\hline\hline
\end{tabularx}
\end{table}

\par
\begin{figure}
     \centering
     \begin{subfigure}[b]{0.49\textwidth}
         \centering
         \includegraphics[width=\textwidth]{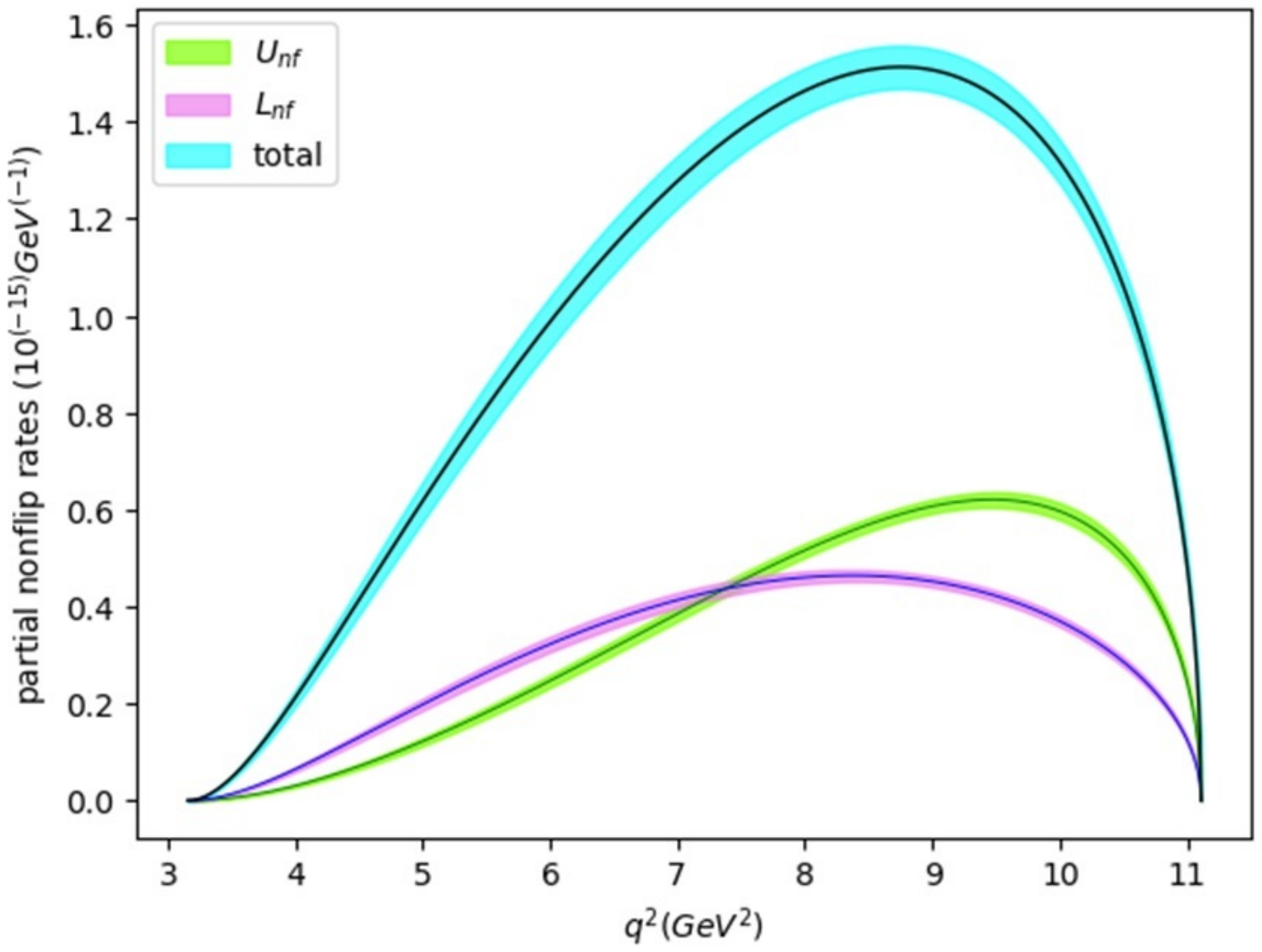}
         
     \end{subfigure}
     \hfill
     \begin{subfigure}[b]{0.49\textwidth}
         \centering
         \includegraphics[width=\textwidth]{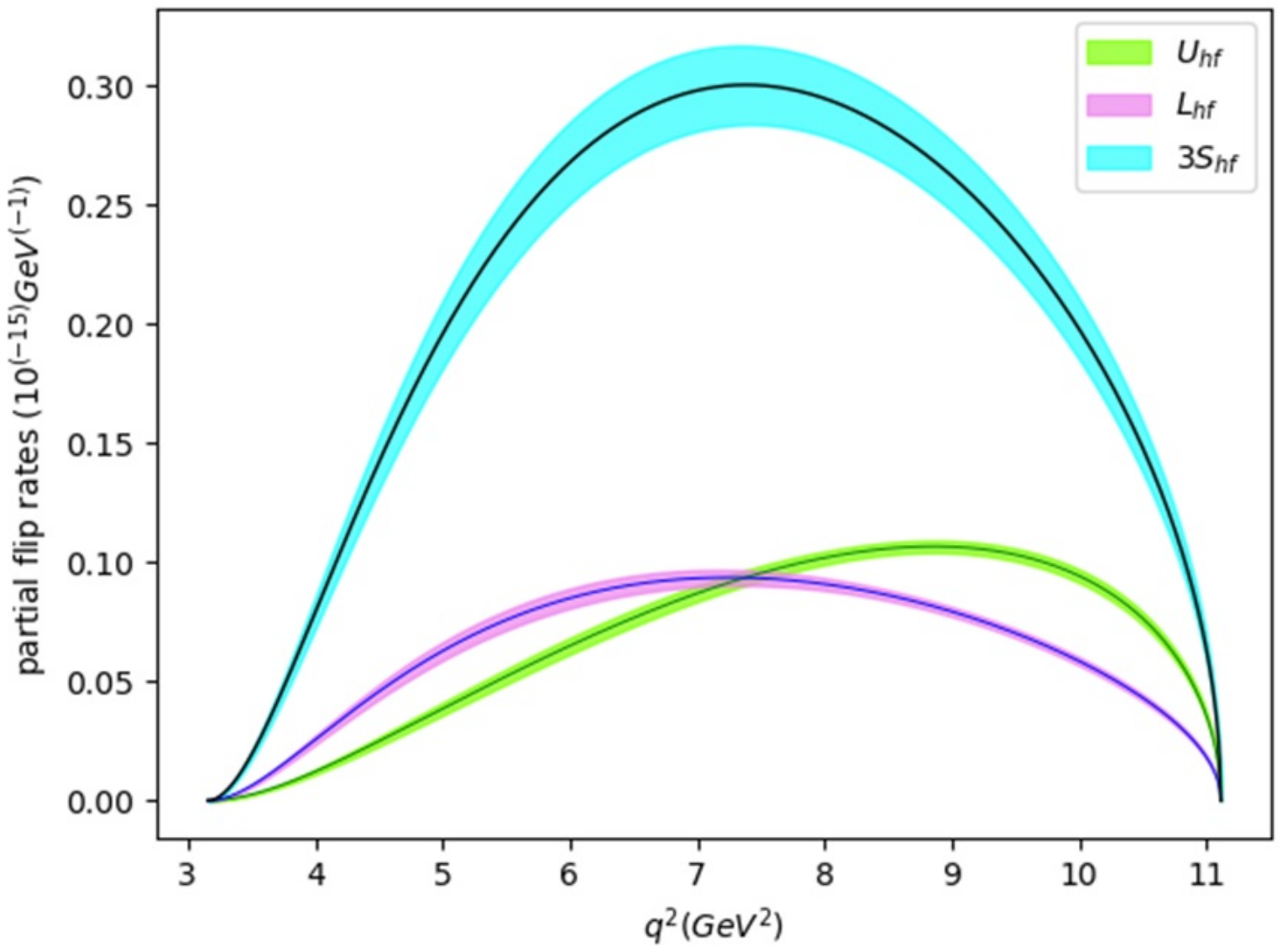} 
     \end{subfigure}
     \caption{Partial nonflip and flip rates versus $q^2$ for the 
    $\tau$ mode.}
    \label{fig: hf_dw}
     \hfill
     
\end{figure}

Our branching fractions for the semi-tauonic channel, $BR(\Lambda _b^0 \to \Lambda _c^ + {\tau ^ - }{\bar \nu _\tau }) = 1.744 \%$ and $1.752 \%$ are in agreement with the recently measurement of LHCb $BR(\Lambda _b^0 \to \Lambda _c^ + {\tau ^ - }{\bar \nu _\tau })= (1.50 \pm 0.16 \pm 0.25 \pm 0.23)\%$ \cite{LHCb:2022piu}. Miao {\it{et. al.}} predicted the branching fractions as follows: $BR(\Lambda _b^0 \to \Lambda _c^ + {e ^ - }{\bar \nu _e }) = 5.81 \times 10^{-2}$, $BR(\Lambda _b^0 \to \Lambda _c^ + {\mu ^ - }{\bar \nu _\mu }) = 5.78 \times 10^{-2}$, and $BR(\Lambda _b^0 \to \Lambda _c^ + {\tau ^ - }{\bar \nu _\tau }) = 1.55 \times 10^{-2}$ in the framework of LCSRs \cite{Miao:2022bga}. Our ratios in Table \ref{Table: BR}, $5.211 \%, 5.195 \%,$ and $1.744 \%$ as well as $5.321 \%, 5.305 \%,$ and $1.752 \%$ are consistent with them respectively. 
The unpolarized total decay width comes from the contributions of longitudinal, transverse, and scalar helicity structure functions. The helicity flip and nonflip partial rates have been obtained by using Eqs. (\ref{Eq: hf}), and (\ref{Eq: nf}) respectively. 
We display the $q^2$ dependence of the partial helicity flip and nonflip rates in Fig. \ref{fig: hf_dw} for the $\tau$ mode using PM configuration. The contributions of helicity flip rates are smaller than the helicity nonflip rates however they play a significant role in the total rates. 
\par
Subsequently, we provide the numerical values for the integrated quantity forward-backward asymmetry $\left\langle {\mathcal {A} _{FB}^\ell } \right\rangle $ in Table \ref{Table: FB asymmetry}, which are obtained by the flip and nonflip rates of Tables \ref{Table: partial decay widths hf} and \ref{Table: partial decay widths nf}. In Ref. \cite{Zhu:2018jet}, the asymmetries of the semileptonic decays for electronic, muonic, and tauonic channels have been reported as -0.03, -0.07, and -0.13 respectively. The magnitude of the asymmetry for the tau lepton process is larger than the two other ones.

\begin{table}[htb]
\caption{Averages of Forward-backward asymmetry.}
\label{Table: FB asymmetry}
\begin{tabularx}{1\textwidth}{>{\centering\arraybackslash}X | >{\centering\arraybackslash}X | >{\centering\arraybackslash}X}

\hline\hline

Mode &  $\left\langle {\mathcal {A} _{FB}^\ell } \right\rangle$ (PM) &  $\left\langle {\mathcal {A} _{FB}^\ell } \right\rangle$ (BGL)
\\
\hline
${e^ - }{{\bar \nu }_e}$ & 
-0.206  &  -0.204
\\
\hline
${\mu^ - }{{\bar \nu }_\mu}$ & 
-0.211  &  -0.209
\\
\hline
${\tau^ - }{{\bar \nu }_\tau}$ & 
-0.334 &  -0.334
\\
\hline\hline
\end{tabularx}
\end{table}

\section{CONCLUSIONS}
\label{section: conclusion}
We have provided a detailed analysis of the semileptonic $\Lambda_b^0 \to \Lambda_c^+$ based on the helicity formalism with the phase-moment parameterization for the form factors in the whole kinematic region. In our analysis, we have included lepton mass effects. Our angular decay formula is unpolarized, and it can be applicable to the corresponding two body baryon decays. We have performed fitting to the LQCD data and obtained the parameters of the form factors, which provided the calculations on the rates. Comparisons between our form factors and those in the LCSR were performed. Our results for the decay rates and their ratios are in agreement with other theoretical approaches and available experimental data. We have also evaluated the leptonic forward-backward asymmetries, which are important for future experiments. In addition to the phase-moment model, we have considered Boyd-Grinstein-Lebed parametrization to find the physical observables for the transition of $\Lambda _b^0 \to \Lambda _c^ + {\ell ^ - }{\bar \nu _\ell }$. As the hadronic transitions of $b \to c$ are widely investigated in the heavy flavour physics experiment, it is beneficial to extend the theoretical study of such modes.

\section*{Acknowlegements}
I would like to thank De-Liang Yao for useful discussions. This work is supported by the Fundamental Research Funds for the Central Universities under Contract No. 531118010379.

\end{document}